\documentclass[aps,pre,twocolumn,superscriptaddress,10pt,%
floatfix]{revtex4-1}
\usepackage{graphicx}
\usepackage{amsmath}
\usepackage{natbib}
\usepackage[utf8]{inputenc}
\usepackage{amssymb,nicefrac}
\usepackage{color}
\usepackage{newfloat,microtype}

\definecolor{cream}{RGB}{222,217,201}
\usepackage{microtype}

\definecolor{cream}{RGB}{222,217,201}
\usepackage{microtype}

\begin{document}

\title{Elastic capsules at
    liquid-liquid interfaces}

\author{Jonas Hegemann}
\affiliation{Physics Department, TU Dortmund University, 44221 Dortmund, Germany}
\author{Horst-Holger Boltz}
\affiliation{Institute for Nonlinear Dynamics, University of G\"ottingen, 37077 G\"ottingen, Germany}
\author{Jan Kierfeld}
\affiliation{Physics Department, TU Dortmund University, 44221 Dortmund, Germany}
\email{jan.kierfeld@tu-dortmund.de}

\begin{abstract}
We investigate the
deformation of elastic microcapsules adsorbed at liquid-liquid
interfaces. An initially spherical elastic capsule at a liquid-liquid
interface undergoes circumferential 
 stretching due to the liquid-liquid surface
tension and becomes lens- or discus-shaped, depending on its bending
rigidity.  The resulting elastic capsule deformation is qualitatively
similar, but distinct from the deformation of a liquid droplet into a liquid
lens at a liquid-liquid interface.
We discuss the deformed shapes of droplets and capsules
adsorbed at liquid-liquid interfaces for a whole range of different surface
elasticities: from droplets (only surface tension) deforming 
into liquid lenses,  droplets with a 
Hookean membrane (finite stretching modulus, zero bending modulus)
deforming into  elastic lenses, to microcapsules (finite stretching and bending
modulus) deforming into rounded elastic lenses. We calculate capsule shapes at
liquid-liquid interfaces numerically using shape equations from nonlinear
elastic shell theory. Finally, we present theoretical results for
the contact angle (or the capsule height) and the maximal capsule 
curvature at the three phase contact line. These results can be used
to infer information about the elastic moduli from optical measurements.
During capsule deformation into a lens-like shape, surface energy of the
liquid-liquid interface is converted into elastic 
energy of the capsule shell giving rise to an overall adsorption 
energy gain by deformation. 
Soft hollow capsules exhibit a
pronounced increase of the adsorption energy as compared to filled
soft particles and, thus, are attractive candidates as foam and 
emulsion stabilizers.
\end{abstract}

\maketitle

\section{Introduction}

Microcapsules, ``hollow microparticles composed of a solid shell surrounding a
core-forming space available to permanently or temporarily entrapped
substances''~\cite{vert2012terminology}, can be produced 
artificially and serve as container and delivery systems 
in many applications, or as a model for
biologically relevant elastic containers, such as cells, virus
capsules and red blood cells
\cite{dinsmore2002colloidosomes,Donath1998,de2010polymeric,orive2004history,  
cook2012microencapsulation,mondal2008phase,martins2014microencapsulation,
aissa2012self,white2001autonomic,gharsallaoui2007applications,
buenemann2008elastic}.
There are various methods to produce artificial microcapsules
\cite{yow2006formation,de2010polymeric} resulting in 
solid shells  comprised of, for example, colloidal particles in 
colloidosomes~\cite{thompson2015colloidosomes,dinsmore2002colloidosomes},
quasi two-dimensional polymerized networks \cite{Rehage2002},
polymer multilayers~\cite{Donath1998,de2010polymeric}, 
and even bacterial films can form elastic capsules~\cite{vaccari2015films}.
Also polyelectrolyte self-assembly in a single microfluidic 
production step has recently been demonstrated \cite{Xie2017}.
The core-forming space can be made from pure liquids,
polymer matrices (gel-like), or solid cores~\cite{gharsallaoui2007applications}.
All these different techniques regarding the solid shell and the core-forming
space give rise to tunable elastic properties and deformation behavior. 

The exact composition of the solid shell and the core-forming space depends on
the specific application, where requirements might be of chemical, biological
or mechanical nature.  Drugs are the main application of microcapsules
nowadays~\cite{martins2014microencapsulation}.  In medicine, microcapsules
enable targeted release of incorporated drugs or cells under certain
conditions \cite{de2010polymeric,orive2004history}.  Moreover, microcapsules
are used for food~\cite{cook2012microencapsulation},
textiles~\cite{mondal2008phase}, cosmetics
\cite{martins2014microencapsulation}, self-healing
materials~\cite{aissa2012self,white2001autonomic,hu2009mechanical} and
powders~\cite{gharsallaoui2007applications}.

For many applications, in particular if rupture and release are 
involved, a characterization of
the mechanical properties of the capsule shell, i.e., its elastic moduli, is
necessary \cite{Vinogradova2006,Neubauer2014}.
Elasticity protects microcapsules from breakage or rupture by converting
externally applied forces into deformation energy. This enables microcapsules
to resist high external loads, pass through thin capillaries, and take diverse
shapes.  

\begin{figure}
 \centering
 \includegraphics[width=\linewidth]{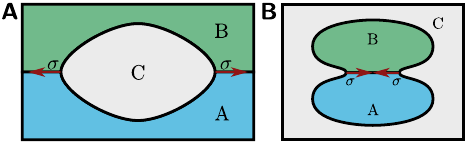}
 \caption
    {
    (A) Capsule stretched into lenticular shape 
   by an interface 
 between two liquid phases A and B outside the capsule. 
  (B) Capsule  compressed into dumbbell shape by 
   an interface between two liquid phases A and B 
   inside the capsule, which can form, for example, 
  by phase separation. 
}
 \label{fig:intro}
\end{figure}

In this paper, we investigate elastic capsules at an interface 
between two liquid phases A and B, see Fig.\ \ref{fig:intro}.
If the capsule adsorbs to an external liquid-liquid interface,
this interface exerts a {\it tensile} line
stress on the capsule as shown in Fig.\ \ref{fig:intro}(A).  
If two liquid phases A and B coexist inside the capsule, 
for example, after a phase separation process, the 
liquid-liquid interfaces exerts a {\em contractile} line tension,
see Fig.\ \ref{fig:intro}(B).
Similar systems with contractile line tensions are 
liquid droplets in contact with a membrane \cite{Kusumaatmaja2011},
two-component vesicles after 
phase separation \cite{julicher1996shape}, or cells during mitosis
\cite{zumdieck2007stress}, where the tension is exerted by the contractile
actin ring. 
We will mainly focus on capsules adsorbed to 
liquid interfaces exerting tensile stresses in this paper, 
but the corresponding shapes for contractile stresses can be obtained 
using the same theoretical approaches provided in this paper.

There have been various studies concerning hard particles at liquid-liquid
interfaces~\cite{park2011janus,park2012configuration} that can be extended
to deformable particles resulting in an additional degree of tunability. 
Recently, the spreading of filled soft particles made from crosslinked
gels (microgel particles) at liquid-liquid interfaces has been investigated
experimentally~\cite{richtering2012responsive}, by molecular dynamics
simulations~\cite{mehrabian2016soft,geisel2015hollow}, and
analytically~\cite{style2015adsorption}. 
When it comes to collective phenomena, experiments with microgel particles 
enclosing solid silica cores have revealed complex packing phenomena at the
interface~\cite{rauh2017compression}. 
Interfaces with hard particles enclosing soft shells also exhibit special 
elasticity with constitutive relations that change 
upon hard core contact~\cite{Knoche2015}.

Soft particles at liquid-liquid interfaces are efficient  
emulsifiers because they can stretch during 
adsorption~\cite{style2015adsorption}.
Adsorption to the liquid-liquid interface takes place if the 
 liquid-liquid surface tension is sufficiently high, such that 
there is a net energy gain from 
a reduction of liquid-liquid interface area. 
Under these conditions, 
soft particles are stretched at the liquid-liquid interface
and  assume an energetically optimal lens-like 
shape~\cite{style2015adsorption,mehrabian2016soft,geisel2015hollow},
which further increases the occupied interface area at the cost of 
increased elastic energy.
The lower this cost, i.e., the softer the particle, the greater the
interface area that gets occupied, and the more the reduction in
the total energy of the liquid–liquid interface by adsorbed
particles.

\begin{figure*}[bth]
 \centering
 \includegraphics[width=0.9\linewidth]{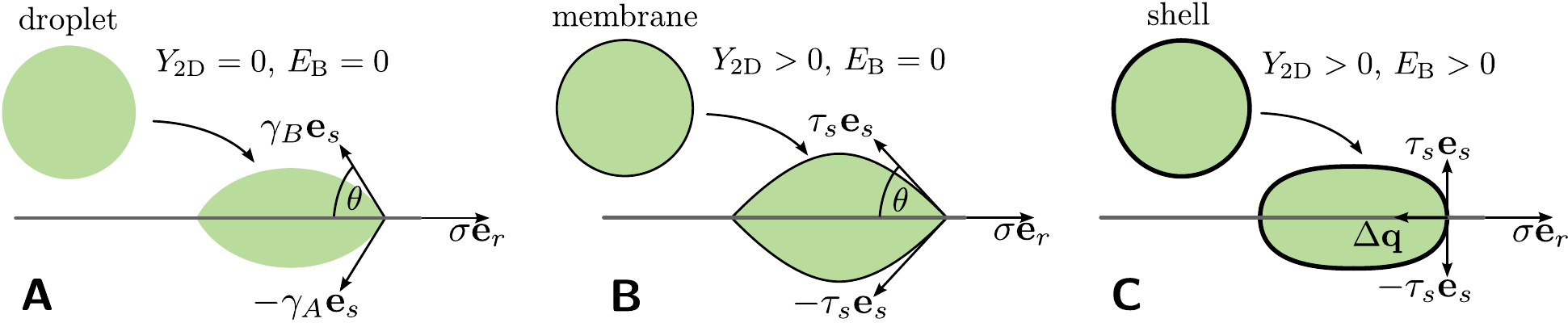}
 \caption{
  Deformation following adsorption to an
  interface between two liquid phases (A and B in the lower and upper 
  half-space, respectively, the sketch depicts 
  the symmetric case $\gamma_\mathrm{A} = \gamma_\mathrm{B}$) 
  for (A) a liquid  droplet, 
  (B) a spherical membrane coated droplet, (C) a spherical 
  shell coated droplet or microcapsule. 
  (A) A liquid droplet deforms into a lenticular
  shape. Such a liquid lens consists of two spherical caps with a kink at the
  interface, where surface tensions are balanced in Neumann's triangle.  
  (B) Adsorption of a thin spherical elastic capsule 
  to a planar liquid-liquid interface yields an
  elastic lens, where the kink at the interface is still preserved but shapes
  are no longer exact spherical caps.  
  Tangential stresses $\tau_s$ and the liquid-liquid 
  surface tension $\sigma$ 
  acting along the liquid interface have to balance at the 
  interface in Neumann's triangle. 
  (C) Finally, adsorbed elastic shells exhibit a lens-like shape with 
  rounded edge due to a finite thickness of the material. 
  Now, tangential stresses $\tau_s$, liquid-liquid 
  surface tension $\sigma$, and 
  the discontinuity $\Delta q$ in the 
  normal transverse shear force density $q$ 
  have to balance each other at the contact line.
}
 \label{fig:num-solB}
\end{figure*}

Therefore, hollow elastic capsules with a thin elastic shell, 
which are much softer than filled particles, are very attractive 
candidates to improve emulsification further.  
We will characterize their deformation behavior in detail in this paper
and show that elastic capsules at a liquid-liquid interface 
take discus- or lens-like shapes due to surface tension in the
interface plane, which leads to an expansion of the capsule circumference
against stretching forces and bending moments in the shell, 
see Fig.\ \ref{fig:num-solB}. The shape of elastic capsules 
resembles the well-known lens-like shapes of a liquid droplet partially
wetting a liquid-liquid interface as shown in 
 Fig.\ \ref{fig:num-solB}(A). Varying the
surface tension versus Young's and bending modulus of the capsule we will 
systematically study the crossover from surface tension dominated 
droplet-like elasticity (liquid lens in Fig.\ \ref{fig:num-solB}(A)) 
to Hookean membrane elasticity for 
finite Young's modulus but zero bending modulus (elastic lens in 
Fig.\ \ref{fig:num-solB}(B)) 
to shell elasticity 
with finite Young's and bending modulus  (rounded elastic lens in 
Fig.\ \ref{fig:num-solB}(C)).

Apart from modifying the adsorption behavior,
capsule deformation by interfacial circumferential 
tension constitutes an
independent hydrostatic method of probing the 
 elastic properties of microcapsules. 
Understanding the deformation under known external
loads allows for elastometry, i.e, the determination of the elastic properties
of the capsule's shell, which in turn can be used to infer information
about the physics and chemistry of the shell as, for example, 
its state of crosslinking.

We will show that the overall shape of the deformed capsule -- 
the height or contact angle of its lens-like shape --
 allows us to infer information 
about the Young's modulus of the capsule shell, whereas the 
meridional curvature at the ``tips'' of the rounded lens, i.e., 
across interfacial plane where the tensile tension acts 
allows us to infer information about the bending modulus of the shell
(if the surface tension of the liquid-liquid interface that 
is exerting the tensile stress is known). 
Other static non-contact 
elastometry methods following the same philosophy are 
the study of deformations of pendant capsules under 
volume changes to obtain elastic moduli
as investigated in Refs.\ \citenum{Knoche2013,Nagel2017,hege2017elastometry},
the study of the edge curvature of buckled shapes to obtain 
the bending modulus  \cite{Knoche2011}, or the study of shapes 
of osmotically buckled capsules to infer the osmotic 
pressure \cite{Knoche2014osmotic}.
Non-contact techniques requiring motion in the surrounding fluid 
are shape analysis in shear flow \cite{chang_olbricht_1993,DeLoubens2016},
in extensional flow \cite{DeLoubens2014},
 and spinning drop rheometry \cite{pieper1998}.

The paper is structured as follows: first, we 
discuss relevant capillary lengths scales and
introduce our elastic 
model for the deformation of a  microcapsule of fixed volume 
 using shape equations.
The external stretching force that  the liquid-liquid interface 
exerts on the capsule is taken into account via 
a Neumann triangle construction for  the 
force equilibrium at the contact line between capsule and liquid-liquid  
interface.
 Then, we discuss numerical solutions for the  resulting shapes 
in three elasticity regimes: the well-known 
droplet limit, where only surface tension acts and the capsule becomes 
equivalent to 
a droplet  partially wetting the liquid-liquid interface, the membrane regime
corresponding to a shell of vanishing thickness but with finite Young's
modulus, and the shell regime, where we also account for bending moments and
transverse shear stress due to the shell's finite thickness. 
We then discuss how to extract Young's and bending modulus 
from characteristics of deformed capsule shapes based on analytical results.
Finally, we use our results on capsule deformation to calculate
the enhancement of the adsorption energy for 
hollow capsules for decreasing shell thickness.

\section{Methods}

In the following, we introduce our model of a microcapsule as a
Hookean shell. We start with 
the elastic model parameters, which also give rise to 
different capillary length scales
controlling the deformation by the liquid-liquid surface tension and the 
relevance of gravitational effects.
We  introduce 
the geometric description for axisymmetric shells, 
the elastic energy,
 constitutive relations, and finally the shape equations, which
can be used to obtain shape profiles by numerical integration.
The shape equation approach has also been used in Refs.\ 
\citenum{Knoche2011,Knoche2013, Knoche2014osmotic,Boltz2015,Boltz2016,
 hege2017elastometry}. Therefore, we defer details to Appendix 
\ref{app:shape}. The deformation by a localized
circumferential stretching force due to the liquid-liquid surface
  tension is the novel aspect in the present problem, which 
is taken into account by employing appropriate matching 
conditions at the three phase contact line with the 
liquid-liquid interface.

\subsection{Elastic parameters and capillary length scales}

The capsule is enclosed by an 
elastic shell of thickness $H$ whose resting shape is a sphere
with radius $R_0$. 
In the limit of a thin shell, $H\ll R_0$, made
from an isotropic and homogeneous elastic material with 
Young's modulus $Y_\mathrm{3D}$, we can use an 
effectively two-dimensional description
with a  two-dimensional Young's modulus $Y_\mathrm{2D} = Y_\mathrm{3D}H$ 
and a bending modulus given by~\cite{Libai1998}
\begin{align}
E_\mathrm{B} &= \frac{Y_\mathrm{2D}H^2}{12(1-\nu_\mathrm{2D}^2)},
\label{eqn:eb}
\end{align}
where 
$\nu_\mathrm{2D}$ is the two-dimensional Poisson ratio. 
Choosing $R_0$ as unit of length and $Y_\mathrm{2D}$ as unit of tension,
the dimensionless bending modulus is given by
\begin{align}
\tilde{E}_\mathrm{B} &= \frac{E_\mathrm{B}}{Y_\mathrm{2D}R_0^2}
      =\frac{1}{\gamma_\mathrm{FvK}} =
    \frac{1}{12(1-\nu_\mathrm{2D}^2)}\frac{H^2}{R_0^2},
\label{eqn:EB}
\end{align}
where $\gamma_\mathrm{FvK}$ is the F\"oppl-von-K\'{a}rm\'{a}n number,
 and the last equality again assumes (cp.\ eq.~\eqref{eqn:eb})
 a thin shell made
from an isotropic and homogeneous elastic material. 
Note that we use the same units throughout the paper, i.e., 
we measure tensions in units of the Young's modulus $Y_\mathrm{2D}$ and lengths 
in units of the resting shape's radius $R_0$.
 Using these natural units we transform 
quantities $x$ to their dimensionless counterparts $\tilde{x}$.
 
By fixing the Poisson ratio $\nu_\mathrm{2D}=1/2$ (corresponding to a linearly
incompressible bulk material) the capsule's elastic response to external
forces is solely determined by the dimensionless bending modulus
$\tilde{E}_\mathrm{B}$. Typical values for
microcapsules range within $\tilde{E}_\mathrm{B}=10^{-10}\dots 10^{-1}$
assuming $E_\mathrm{B}=10^{-16}\dots10^{-14}\,\mathrm{Nm}$,
$Y_\mathrm{2D}=10^{-2}\dots 10^{0} \,\mathrm{N/m}$ and $R_0=10^{-6}\dots
10^{-3}\,\mathrm{m}$ \cite{hege2017elastometry}. 
If $\tilde{E}_\mathrm{B} \sim 10$ (corresponding to $H\sim R_0$ 
for $\nu_\mathrm{2D}=1/2$) 
we expect to obtain (at least qualitatively) the crossover to the limit of a 
filled soft particle that has been considered in Refs.\ 
 \citenum{mehrabian2016soft,geisel2015hollow,style2015adsorption}.

Based on Young's modulus we can estimate the 
typical deformation.
The capsule will be stretched  by 
the surface tension $\sigma$ of the liquid-liquid interface, which 
acts along the circumference of the capsule. 
A deformation by $\Delta R$ at the liquid-liquid interface
causes strains $\sim \Delta R/R_0$ and  
costs an elastic stretching energy $E_{\rm el} \sim Y_\mathrm{2D} R_0^2(\Delta
R/R_0)^2 \sim Y_\mathrm{3D} H R_0^2(\Delta R/R_0)^2$ but gains 
an interfacial energy $E_\sigma \sim \sigma R_0 \Delta R$. 
The resulting strain  is of order 
$\Delta R/R_0 \sim \sigma/Y_\mathrm{2D}\sim 
\sigma/Y_\mathrm{3D}H$, i.e., {\it independent}
of capsule size but limited by shell thickness for 
{\it hollow} capsules. 
The behavior is different for  a 
{\it filled} soft particle, where we 
expect $\Delta R/R_0 \sim \sigma/Y_\mathrm{3D}R_0$, i.e., 
small filled particles deform stronger than large filled particles.
For filled soft particles, 
this allows to define an {\it elastocapillary length} 
$L_{\sigma} = \sigma/Y_\mathrm{3D}$ 
such that deformations by the  surface tension 
$\sigma$ become large  for small particles 
 $R_0 \ll L_{\sigma}$  \cite{style2015adsorption,Vella2015}.
For hollow capsules with a soft shell, on the other hand, 
deformations become  large if the 
shell thickness $H$ is sufficiently small compared 
to the elastocapillary length: $H\ll L_{\sigma}$ (or $\sigma \gg Y_\mathrm{2D}$).
We can generally state that hollow capsules deform more significantly
than filled soft particles of the {\it same} size.

Although  the size of the capsule is not relevant 
for deformation by  the liquid-liquid interface, 
it will play a role for deformation by
gravitational forces.
The typical gravitational energy gain upon deformation by 
$\Delta R$ is $E_g \sim \Delta \rho g R_0^3 \Delta R$, where 
$\Delta \rho$ is the density difference between the liquids inside
and outside the capsule. Gravitational energy competes with the 
elastic deformation energy  $E_{\rm el}$ resulting in strains 
$\Delta R/R_0 \sim  \Delta \rho g R_0^2/Y_\mathrm{2D}$, i.e., 
deformation by gravitation is relevant for capsules larger than a 
{\it gravitational  elastocapillary length}
 $L_g = (Y_\mathrm{2D}/\Delta \rho g)^{1/2}$. 

Moreover, capsule size is also relevant for the  deformation  of the 
liquid-liquid interface by the gravitational force on the capsule, 
i.e., the shape of the meniscus. The balance of the 
gravitational energy $E_g$  and the interfacial energy $E_\sigma$ 
shows that the meniscus is curved on the scale 
of the  {\it capillary length}
 $L_c = (\sigma/\Delta \rho g)^{1/2}$.
Only capsules larger than $L_c$ give rise to a relevant 
 curvature of the  liquid-liquid interface \cite{Vella2015}.

We will focus on soft hollow capsules with 
 $H\ll R_0$,  for which gravitational 
effects both for the meniscus and
 capsule deformation are negligible, i.e., $R_0 \ll L_g,L_c$,
and strains of the capsule do not become large but are 
non-negligible, i.e., 
$H \ge   L_\sigma$ or $\sigma \le Y_\mathrm{2D}$. 
For artificial  microcapsules  this is a generic situation, 
as the following estimates show:
For microcapsules with $R_0 \sim 10 \,\mathrm{\mu m}$
with a  typical soft capsule shell material with 
$Y_\mathrm{2D}\sim 10^{-2}...10^{-1} \,\mathrm{N/m}$ and a 
shell thickness $H\sim 0.1\, \mathrm{\mu m}$ (corresponding to 
$Y_\mathrm{3D}=Y_\mathrm{2D}/H \sim 10^5...10^6 \,\mathrm{Pa}$), 
with a  liquid-liquid 
interfacial tension (e.g.\ oil-water) of 
$\sigma \sim  5\cdot 10^{-2} \,\mathrm{N/m}$, and with 
$\Delta \rho \sim  10^2\, \mathrm{kg/m^3}$, we find 
a capillary length $L_c \sim 7 \,\mathrm{mm}$ and a   
gravitational  elastocapillary length $L_g \sim 3...10 \,\mathrm{mm}$, 
which are much larger than $R_0$, 
and an elastocapillary length 
$L_\sigma \sim 5\cdot 10^{-2}...10^{-1}\,\mathrm{\mu  m}$, 
which is of the order of the thickness $H$.

\subsection{Parametrization}

We consider the axisymmetric elastic shell as a surface of revolution 
around the $z$-axis (see Fig.\ \ref{fig:interface-patch} in the Appendix
for details).
The shell contour (its generatrix) is given in 
cylindrical coordinates $(r(s_0),z(s_0))$,
where $s_0$ is the arc length of the undeformed
shape and $r$ is the distance from the $z$-axis. 
The total arc length of the contour is $L_0$, i.e.,  $s_0\in[0,L_0]$. 
 The arc length element of
the deformed shape derives as $\mathrm{d}s = \sqrt{r'(s_0)^2 +
  z'(s_0)^2}\,\mathrm{d}s_0$, and the unit tangent vector $\vec{e}_s =
(\cos\psi,\sin\psi)$ ($\psi$ being the angle between $\vec{e}_s$ and the
$r$-axis) gives the orientation of a capsule patch relative to the axis of
symmetry.  
The undeformed reference shape shall be given by a
sphere with rest radius $R_0$, i.e., $(r_0(s_0),z_0(s_0))=
(R_0\sin(\pi s_0 / L_0), z_0+R_0(1-\cos(\pi s_0 / L_0)))$, 
which generates (by revolution around the $z$-axis)
a sphere with radius $R_0$ whose lower apex (intersection 
with the symmetry axis) is located at $z_0$ and $s_0=0$.

\subsection{Elastic energy}

Stretching deformations with respect to the undeformed spherical shape can be
expressed in terms of the stretches
$\lambda_s = \mathrm{d}s/\mathrm{d}s_0$ and $\lambda_\phi = r/r_0$, and
bending deformations in terms of the principal curvatures 
$\kappa_s = \mathrm{d}\psi/\mathrm{d}s$ and
$\kappa_\phi = \sin\psi / r$, which derive from the second fundamental form of
a surface of revolution~\cite{docarmo}.  
The deformation energy is formulated best in terms of the (stretching) strains 
$e_{s,\phi} = (\lambda_{s,\phi}^2-1)/2 \approx \lambda_{s,\phi} - 1$
employing a small strain approximation and bending strains 
$K_{s,\phi} = \lambda_{s,\phi}\kappa_{s,\phi}-\kappa_{s_0,\phi_0}$, 
where we expanded up to linear order in deviations from the 
undeformed reference shape with $\lambda_s=\lambda_\phi=1$ and 
$\kappa_{s,\phi}=\kappa_{s_0,\phi_0}$. 
\footnote{
Systematic derivations starting from thin three-dimensional 
elastic materials \cite{ciarlet2006,Efrati2009} show that 
factors $\lambda_{s,\phi}$ should be contained 
in the definition of bending strains.
 As a result, a spherical shell that is 
uniformly stretched  from radius $R_0$ to radius $R$ has 
$\lambda_{s,\phi} = R/R_0$ and $\kappa_{s,\phi}=1/R \neq
\kappa_{s_0,\phi_0}=1/R_0$
but vanishing $K_{s,\phi}=0$, which is reasonable because 
it has obviously not developed  bending moments but only 
stretching tensions. 
For strictly two-dimensional materials with 
bending energy functionals such 
as the Helfrich energy, the curvature cannot be changed by 
stretching the material.
}
We consider   hyperelastic materials,
 whose elastic energy can be expressed in terms of a local
energy density, and use a Hookean surface energy density
\cite{Libai1998}
\begin{align}
\begin{split}
w(s_0) \mathrm{d}A_0 &= \frac{1}{2} 
\bigg( \frac{Y_\mathrm{2D}}{(1-\nu_\mathrm{2D}^2)}
    (e_s^2+2\nu_\mathrm{2D}e_se_\phi+e_\phi^2) \\
&~~~~~+ E_\mathrm{B}(K_s^2+2\nu_\mathrm{2D}K_sK_\phi+K_\phi^2) 
+\lambda_s\lambda_\phi\gamma\bigg) \mathrm{d}A_0
\end{split}
\label{eqn:hooke-energy}
\end{align}
with the linear approximation 
$e_{s,\phi} \approx \lambda_{s,\phi} - 1$.
The three terms in the energy density correspond to the three contributions
from stretching, from bending, and from an effective 
interfacial tension between the fluids outside and inside the capsule.
We consider capsules smaller than the gravitational elastocapillary length
$L_g$, such we can neglect gravitational body forces in the energy.
We explicitly
state the undeformed surface element $\mathrm{d}A_0$ to highlight the fact
that this energy functional operates on the undeformed surface which is
important for computing stresses from it.  

The energy (\ref{eqn:hooke-energy})
explicitly contains a contribution
from an isotropic effective surface tension $\gamma$
between the outer liquids and the capsule.
Such a contribution arises 
either as the sum of surface tensions of the liquid outside 
with the outer capsule surface and the liquid inside with the 
inner capsule surface or, if
the capsule shell is porous such that there is still contact 
between the liquids outside and inside the capsule,  
with additional contributions 
from the surface tension between outside and inside liquids. 
In general, the liquid phases A in the lower half-space and B 
in the upper half-space 
will have different surface tensions $\gamma=\gamma_\mathrm{A}$ and 
$\gamma=\gamma_\mathrm{B}$ with the capsule. 
 We will distinguish between the simpler {\it symmetric} case
 $\gamma_\mathrm{A} = \gamma_\mathrm{B}$ (which is also 
depicted in Fig.\ \ref{fig:num-solB}) and the 
general  {\it asymmetric} case with $\gamma_\mathrm{A} \neq
\gamma_\mathrm{B}$.
We will mainly focus  on the symmetric case throughout 
the paper, where the AB-interface acts along the equator
of the capsule.

The {\it dimensionless surface tension} $\gamma/Y_\mathrm{2D}$ 
(with $\gamma\equiv \gamma_\mathrm{A} = \gamma_\mathrm{B}$ in the 
symmetric case)
 is another important control parameter 
of the capsule, which governs the crossover from a liquid 
droplet to an elastic Hooke membrane or shell: 
For $\gamma/Y_\mathrm{2D}\gg 1$ the capsule behaves as a liquid droplet
and assumes lens-like shapes as in  Fig.\ \ref{fig:num-solB}(A).
For $\gamma/Y_\mathrm{2D}\ll 1$  the capsule behaves either 
as a Hookean membrane also assuming a lens shape as  in
Fig.\ \ref{fig:num-solB}(B) (for small bending moduli $\tilde{E}_\mathrm{B}$) 
or as an elastic shell assuming a 
rounded lens shape as in Fig.\ \ref{fig:num-solB}(C)
(for larger bending moduli $\tilde{E}_\mathrm{B}$).

We assume a spherical rest shape of the capsule, i.e., the capsule 
is {\it not} synthesized  at the liquid-liquid interface but inside 
one of the liquid phases A or B {\it before} it is adsorbed to the
interface. 
As already worked out in Ref.\ \citenum{style2015adsorption},
adsorption of a spherical particle  to the AB-interface also depends
on the surface tension $\sigma$ of the AB-interface: 
If   $\gamma_\mathrm{A} > \gamma_\mathrm{B}+\sigma$  
the particle stays in liquid B,
if $\gamma_\mathrm{B} > \gamma_\mathrm{A}+ \sigma$ the particle stays in liquid A.
For sufficiently large $\sigma > |\gamma_\mathrm{A} - \gamma_\mathrm{B}|$, 
the particle will always adsorb to the interface.

We note that the energy functional (\ref{eqn:hooke-energy})
(with $\nu_\mathrm{2D}=1/2$) also correctly describes the low-strain behavior
of more sophisticated energy functionals, e.g., of the Mooney-Rivlin
type~\cite{Libai1998,Barthes-Biesel2002}. 
In this regard, our model as well as the constitutive
relations given below are a small strain limit.
The Hookean small strain description is often sufficient and 
the most simple  though non-trivial choice. 
For larger strains, Mooney-Rivlin, Skalak or other 
elastic energies are more appropriate. 
In general, we expect large strains  to occur for 
stretching tensions $\sigma > Y_\mathrm{2D}$ ($H <   L_\sigma$)
and $\sigma > \gamma$.
Tensile stresses  $\sigma \le Y_\mathrm{2D}$  are realistic 
for typical experimental situations. 
In Appendix 
\ref{app:shape_equations},  we find that,  for $\sigma \le Y_\mathrm{2D}$,
meridional strains remain small, whereas circumferential stretching 
factors can become locally  large, where the liquid-liquid tension acts
 (reaching values  $\lambda_\phi \simeq 1.8$
for  $\sigma = Y_\mathrm{2D}$, see Fig.\ \ref{fig:lambdas} in the Appendix).
We do not expect, however, that any of our results 
will qualitatively change if  nonlinear elastic laws  beyond 
Hookean elasticity are used in this regime.  
This is also what has been 
found for deflated 
 pending capsule shapes in Ref.\ \citenum{hege2017elastometry}.

\subsection{Constitutive relations}

Variation  of the elastic energy 
 with respect to the strains $e_{s,\phi}$ gives the
tensions $\tau_{s,\phi}$; variation   with respect to the curvatures
$K_{s,\phi}$ gives the bending moments $m_{s,\phi}$. 
This gives the corresponding 
 {\em constitutive relations} of the capsule material for 
a Hookean elastic material, 
\begin{align}
\begin{split}
\tau_{s,\phi} &= \frac{1}{\lambda_{\phi,s}}\frac{\partial w}{\partial e_{s,\phi}}
=\frac{Y_\mathrm{2D}}{1-\nu_\mathrm{2D}^2}
  \frac{1}{\lambda_{\phi,s}}(e_{s,\phi}+\nu_\mathrm{2D} e_{\phi ,s}) + \gamma, 
\\
m_{s,\phi} &= \frac{1}{\lambda_{\phi,s}}\frac{\partial w}{\partial K_{s,\phi}} 
=\frac{E_\mathrm{B}}{\lambda_{\phi,s}}(K_{s,\phi} +\nu_\mathrm{2D}K_{\phi,s}),
\end{split}
\label{eqn:constitutive-laws}
\end{align}
which are nonlinear since the Cauchy stresses are defined with respect to the
deformed arc length, but the surface energy density measures lengths in terms
of the undeformed arc length.  Note that the surface tension $\gamma$ gives a
constant and isotropic contribution to the tensions $\tau_s$ and $\tau_\phi$.

\subsection{Shape equations}

The equilibrium shape of an infinitesimal thin shell is described by local
stress equilibrium in 
tangential and normal direction; elastic shells of
finite thickness additionally require torque (bending moment) balance
(see Appendix \ref{app:shape_equations}).  
In
combination with the constitutive 
laws \eqref{eqn:constitutive-laws} and three
differential equations following from cylindrical parametrization 
the stress and moment equilibrium 
lead to  a closed  system of six shape equations 
for axisymmetric  Hookean shells,
\begin{align}
\begin{split}
r'(s_0) &{=} \lambda_s \cos\psi, \hspace{2mm} 
 z'(s_0) {=} \lambda_s \sin\psi, \hspace{2mm} 
 \psi'(s_0) {=} \lambda_s \kappa_s, \\
\tau_s'(s_0) &= \lambda_s\left(\frac{\tau_\phi -\tau_s}{r}
 \,\cos\psi +\kappa_s q + p_s\right), \\
m_s'(s_0) &= \lambda_s \left(\frac{m_\phi - m_s}{r}\,\cos\psi - q\right),\\
q'(s_0) &= \lambda_s\left(-\kappa_s \tau_s 
           -\kappa_\phi\tau_\phi -\frac{q}{r}\,\cos\psi + p\right),
\end{split}
\label{eqn:shape-equations}
\end{align}
where the quantity $q$ is the transverse shear stress,
 $p = p_0 + p_n$ is the total internal normal pressure, 
and $p_s$ is the shear pressure. 
Since we consider a closed microcapsule 
encapsulating an incompressible liquid phase, $p$ will have a hydrostatic 
contribution, $p_0$, that has to be fixed by the  volume constraint
(see Appendix \ref{app:shape_equations}).
Throughout the paper we consider closed microcapsules 
containing  an incompressible liquid phase. Therefore, 
all  deformations are at  fixed volume, which 
is given by $V=4\pi R_0^3/3$  for an elastic capsule with 
spherical rest shape of radius $R_0$. 
External forces from the surface tension $\sigma$ of the AB-interface 
enter via the additional 
normal  and shear pressures $p_n$ and $p_s$ and will be discussed 
in the following section in detail. 
We consider capsules smaller than the gravitational elastocapillary length
$L_g$, such that we can 
 neglect a gravitational contribution  $-\Delta \rho g z$ to the pressure $p$ 
in the last shape equation,  see also section
\ref{sec:grav}  below.

The first three equations in  \eqref{eqn:shape-equations}
 are geometric relations.
The fourth and sixth equations in  \eqref{eqn:shape-equations}
describe the tangential and normal force equilibrium, respectively. 
The fifth equation is the equilibrium of bending moments.  
Equivalently,  the fourth and fifth   shape equations 
in  \eqref{eqn:shape-equations} can be obtained by 
a variational approach, where we 
 minimize the total free energy 
\begin{align}
  G&= \int w(s_0) \mathrm{d}A_0 - p_0V +E_\sigma
  \label{eqn:Gvar}
\end{align}
  with respect to
the independent functions $r(s_0)$ and $\psi(s_0)$ 
(we repeat this calculation from  Ref.\ \citenum{Knoche2011}
in Appendix \ref{app:variational} for completeness of the presentation)
and where $E_\sigma$ is the potential energy for the 
external forces from the surface tension $\sigma$ (see following section). 
The sixth shape equation in   \eqref{eqn:shape-equations}
is equivalent to  an additional algebraic relation 
$q= -\tau_s \tan\psi + p r/2\cos\psi$ (see also  eq.\
(\ref{eqn:q}) in Appendix \ref{app:match}), which is obtained 
in this variational calculation.
The shape equations \eqref{eqn:shape-equations} are
closed by eliminating $\lambda_s$ and $\tau_\phi$ by using the two 
constitutive relations for stresses and strains 
from eq.\  \eqref{eqn:constitutive-laws},
using the geometric relation $\kappa_\phi = \sin\psi / r$, 
and eliminating 
$\kappa_s$ and $m_\phi$ by using the two 
constitutive relations for bending moments and bending strains 
from eq.\  \eqref{eqn:constitutive-laws}.
This procedure is explained in detail in Ref.\ \citenum{Knoche2011}.

The shape equations in the form 
(\ref{eqn:shape-equations}) are still  independent 
of the elastic material law and only contain stress and moment 
equilibrium and 
geometrical relations but no information on the 
elastic material law, i.e., the constitutive 
relation. The constitutive relation is needed 
to close the shape equations. 
We use the nonlinear Hookean elasticity 
 \eqref{eqn:constitutive-laws} but also other constitutive 
relations can be implemented. In Ref.\ \citenum{hege2017elastometry}
it has been explicitly shown how to use a Mooney-Rivlin relation 
to close the shape equations (\ref{eqn:shape-equations}).

Because the six shape equations  \eqref{eqn:shape-equations} are 
of first order, six boundary conditions are needed. 
Boundary conditions at the apices ($s_0=0$ and $s_0=L_0$) are 
$r(0)=r(L_0)=0$ because the capsule is closed, 
$\psi(0)=\pi-\psi(L_0)=0$ because there are no kinks, and 
 $q(0)=q(L_0)=0$  because there are no point loads at the apices which 
could cause a transverse shear stress $q$. 
The boundary conditions to the remaining
 quantities $z$, $\tau_s$, and $m_s$  are a
priori unknown,  and we have to solve the 
shape equations \eqref{eqn:shape-equations}
 by a shooting method
as explained in the Appendix \ref{app:shooting} to 
fulfill boundary conditions at both apices.

\subsection{Matching conditions at the liquid-liquid interface}
\label{sec:eqstress}

The aim of this paper is to study the
deformation behavior of microcapsules 
during adsorption at a planar liquid-liquid
interface, which can be found, for example, 
between two horizontally layered immiscible liquids A and B. 
We assume the AB-interface to be
in the horizontal plane at $z=0$ and define
the corresponding arc length $s_0 =\ell$ by $z(\ell)=0$.
Gravity can lead to a lowering  of the adsorbed capsule, which will 
deform the AB-interface \cite{Vella2015}.  
We focus on capsules smaller than the capillary length 
 $R_0\ll L_c$ such that this effect can be neglected.
 We will only briefly 
discuss gravity effects in more detail  in the next section
\ref{sec:grav}.
Neglecting gravity we have a  purely  horizontal 
stretching force from the interfacial tension $\sigma$ from 
the AB-interface. We will refer to $\sigma$  as the {\em interface load}
because  $\sigma$ is responsible for stretching the 
capsule.

Adsorption of an object that forms a circular cross-section
with the horizontal interface changes the interfacial surface energy by 
$E_\sigma = -\sigma \pi r^2(\ell)$.
From this energy  we derive the  force density
from the interface load 
by variation with respect to shape changes,
\begin{equation}
\vec{f}_\sigma(s_0)  = \sigma\delta(s_0-\ell)\vec{e}_r,
\label{eqn:fsigma}
\end{equation}
which  acts normal to  the contact line of the capsule at $z=0$
($z(s_0=\ell)=0$)
in horizontal direction.  
The  force density  \eqref{eqn:fsigma} 
can be formally 
incorporated into the shape equations \eqref{eqn:shape-equations}
as additional normal and shear pressure contributions
$p_n$ and $p_s$
(see eq.\ (\ref{eqn:ext-forces}) in Appendix \ref{app:shape_equations}),
which cannot be integrated over directly as they are singular.
Instead, we can also solve \eqref{eqn:shape-equations}  piecewise 
for $z>0$ and $z<0$ (corresponding to $s_0<\ell$ and $s_0 >\ell$),
and derive  matching conditions at the three phase contact line 
at $z=0$ or $s_0=\ell$ 
to account for the force density \eqref{eqn:fsigma}.

Because the shape equations \eqref{eqn:shape-equations} are 
first order, 
the number of matching conditions has to equal the 
 number $n$ of 
 shape equations \eqref{eqn:shape-equations} if the 
interface position $s=\ell$ along the capsule  is known. 
In the symmetric case $\gamma_\mathrm{A}=\gamma_\mathrm{B}$, upper and lower 
part of the capsule are related by reflection symmetry. 
Then, the AB-interface is located at the known arc length $s_0=L_0/2$,
such that $n$ matching conditions are required. 
In the general asymmetric case  the AB-interface,  also  $\ell$ has to 
be determined, such that $n+1$ matching conditions are required. 

We first discuss the symmetric case.
For a Hookean shell, we have the full set of 
$n=6$ shape equations \eqref{eqn:shape-equations}.
 The shell becomes an elastic  membrane in  the limit of vanishing 
$H\approx 0$,
i.e., vanishing bending modulus  $E_\mathrm{B}\approx 0$,
 which implies vanishing bending moments $m_s = 0$ and 
vanishing transverse shear stress $q = 0$, such that only $n=4$ 
shape equations are left. A liquid droplet cannot support 
elastic stresses, such that $Y_\mathrm{2D}\approx 0$, there 
is no reference shape, such that $\lambda_s=\lambda_\phi=1$, and 
also  tangential stresses $\tau_s=\tau_\phi=\gamma$ can 
be eliminated from
\eqref{eqn:shape-equations}, leading to 
 $n=3$ equations equivalent to the well-known Laplace-Young shape equations,
see eqs.\ \eqref{eqn:laplace-young-se} in the Appendix.

In the symmetric case, we need 
$n$ matching conditions  to fit piecewise solutions for the upper and lower
half uniquely together. 
First, we have to impose  continuity
for the  variables $r$ and $z$, which ensures a  
closed capsule  shape. In addition, we impose $z=0$ at the interface
resulting in three matching conditions.

The remaining matching conditions are derived in the Appendix 
  \ref{app:match} from variation of the total free energy
(\ref{eqn:Gvar}).
We find two matching conditions \eqref{eqn:q_match}
and \eqref{eqn:Young}
corresponding to  the balance of interfacial forces 
in $z$- and $r$-direction
at the three phase contact line, i.e., a Neumann triangle 
construction \cite{neumann1894} (see also 
Fig.\ \ref{fig:num-solB}).
Only force balance in $r$-direction involves 
the interfacial force \eqref{eqn:fsigma} in the absence of gravity 
effects. 
In the variational calculus the force balance equation in $r$-direction 
is obtained as a Weierstrass-Erdmann condition from 
variation with respect to $r$. 
In $z$-direction, 
the interfacial forces from the upper an lower part of the capsule 
 not only balance 
each other at the AB-interface 
in the Neumann construction, 
but each of them is also balanced by the  pressure force  
$p_0 r/2$ because both the upper and the lower 
part of the capsule must be force-free in $z$-direction in equilibrium
(see eq.\ (\ref{eqn:U}) in the Appendix). 
Therefore, continuity of  the  variable $r$ is actually {\it equivalent}
to the Neumann matching condition  \eqref{eqn:q_match}
in $z$-direction. If both conditions are used, the set of matching 
conditions becomes over-determined. Interestingly, using both equations in an 
over-determined set of matching conditions makes our numerical analysis 
typically more stable.

For the Hookean membrane, two continuity conditions for 
$r$ and $z$,  the value $z=0$  at the AB-interface,  and 
the   Neumann triangle  condition \eqref{eqn:Young} in $r$-direction, 
are the required  $n=4$ matching conditions to make the problem 
well-posed. 
For the Hookean shell we also get a continuity condition for $\psi$ 
ensuring a smooth shape without kinks (which are forbidden
by bending energy) and
obtain an additional  moment equilibrium \eqref{eqn:m_cont}  at the   
contact line (as a Weierstrass-Erdmann condition 
from variation with respect to  $\psi$)
resulting in the required  $n=6$ conditions 
for the Hookean shell.

In the general asymmetric case,  the AB-interface 
is located at an arc length $s_0=\ell$ ($z(\ell)=0$), 
which has also to be determined from variation of the total free energy. 
In Appendix \ref{app:match} we find an  additional transversality condition
\eqref{eqn:transversality} by equating the boundary terms
 $\propto \delta \ell$
in the variational calculus. This condition is equivalent to the statement,
that the discontinuity of the surface energy density $w$ originates only
from  the discontinuity of the surface tension with the 
surrounding liquid $\gamma$, which jumps from 
$\gamma_\mathrm{A}$ to $\gamma_\mathrm{B}$ at the AB-interface.
This, in turn, is  equivalent to requiring that the 
 elastic contributions to the meridional tension $\tau_s-\gamma$ 
are continuous at the AB-interface (i.e., their absolute values 
are continuous, their directions can be discontinuous, for example, 
in the Hookean membrane case in an elastic lens, 
see Fig.\ \ref{fig:num-solB}(B)).

\subsection{Gravity}
\label{sec:grav}

\begin{figure}[htb]
\centering
\includegraphics[width=\linewidth]{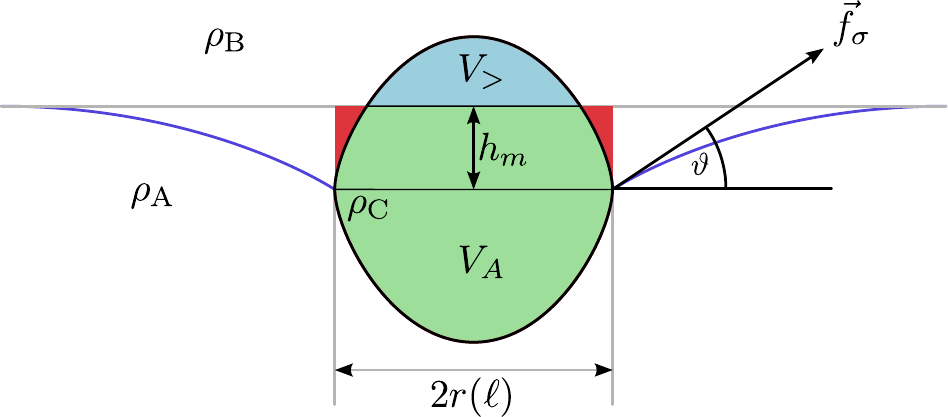}
\caption{Elastic capsule subjected to gravitational forces 
  at a liquid-liquid interface. The total body force $\vec{f}_\mathrm{body}$
  (see eq.\ \protect\eqref{eq:fbody}) 
  acting on the capsule volume is compensated by the vertical 
  component of the interface load $\vec{f}_\sigma$ (see 
  eq.\ \protect\eqref{eqn:fsigma}) (integrated along
  the circular three phase contact line).}
\label{fig:curved}
\end{figure}

Body forces, such as gravitation and buoyancy, that act upon the capsule
can  bend the AB-interface \cite{Vella2015} and 
can deform the capsule. 
We already showed that gravity leads to a curved meniscus 
of the AB-interface if the capsule radius $R_0$ is larger than 
the capillary length $L_c = (\sigma/\Delta \rho g)^{1/2}$
and that gravity deforms the capsule if the capsule radius $R_0$ is 
larger than the  gravitational  elastocapillary length
 $L_g = (Y_\mathrm{2D}/\Delta \rho g)^{1/2}$. Both capillary lengths 
are typically of the order of millimeters, such that gravity 
can be neglected for microcapsules. 
Now we want to briefly 
address in more detail, how our model had to be modified 
in order to include all gravity effects.
The involved quantities are explained in Fig.\ \ref{fig:curved}.

For an axisymmetric capsule the total body force containing gravitational 
and buoyancy force is directed in $z$-direction. 
The AB-interface contacts the capsule at $s_0=\ell$  but if this 
interface bends downward we no longer have $z(\ell)=0$ as without gravity 
but $z(\ell) = -h_m$ with a height $h_m>0$ of the meniscus (if $z=0$ is 
the position of the AB-interface at infinity, see below).  
One contribution to  the buoyancy force is given by the weight difference 
of the volume $\pi r(\ell)^2 h_m- (V_B-V_>)$
 (red area in Fig.\ \ref{fig:curved})
of the liquid A that has been
displaced by liquid B vertically above the contact line 
and the capsule surface up 
to the original horizontal AB-interface at $z=0$  \cite{Keller1998}, 
where $V_>$ is the capsule volume above the  $z=0$ plane
(blue area in Fig.~\ref{fig:curved}; $V_>=0$  if the entire capsule 
is  below the  $z=0$ plane).
The other contribution to the buoyancy force is given by the weight 
difference of  volumes $V_A+V_B-V_>$ (green area in Fig.\ \ref{fig:curved}) 
of liquid A  
and $V_>$ (blue area)  of liquid B  
that have been displaced by the capsule interior C.
Both contributions add up to
\begin{equation}
   \vec{f}_\mathrm{body} = \left[g (\rho_\mathrm{A}- 2\rho_\mathrm{C}) V 
     +g(\rho_\mathrm{A}-\rho_\mathrm{B})
      (\pi r(\ell)^2 h_m-V_B) \right] \vec{e}_z.
\label{eq:fbody}
\end{equation}

If $h_m>0$ and the AB-interface bends, the 
interface load  $\vec{f}_\sigma$, see eq.\ (\ref{eqn:fsigma}),  also has  a
component in $\vec{e}_z$ direction, which, integrated along the circular
cross-section, exactly compensates $\vec{f}_\mathrm{body}$ in equilibrium,
\begin{align}
\vec{f}_\mathrm{body} + \sigma \sin(\vartheta)2\pi r(\ell) \vec{e}_z
   &= 0\,\text{.}
\label{eq:theta}
\end{align}
This determines 
the angle  $\vartheta$  between the interface tangent and the $r$-axis 
at the three phase contact line.

If $\vartheta$ is known, the complete 
shape $z_\mathrm{AB}(x,y)$ of the AB-interface 
follows from the Laplace equation,
which can be written in the form 
\begin{equation}
  -\vec{\nabla} \cdot \vec{n} = \frac{z_{\mathrm{AB}}}{L_c^2}
\label{eq:zAB}
\end{equation}
with a capillary length $L_c = (\sigma/\Delta \rho_\mathrm{AB} g)^{1/2}$,
which is determined by the density difference 
$\Delta \rho_\mathrm{AB}=\rho_\mathrm{A} -\rho_\mathrm{B}$ of the 
two liquid phases, and with the unit normal vector $\vec{n}$ onto 
the AB-interface \cite{Finn86}.
We consider the case of equal pressures in phase A and B
in eq.\ (\ref{eq:zAB}) such that the interface becomes planar 
at infinity; we choose the $z$-coordinates such that 
the AB-interface is at $z=0$ at infinity.

Moreover, the shape of the capsule changes in the presence of gravity.
First, gravity 
  gives rise to a hydrostatic pressure contribution 
$\Delta \rho g z$ to the pressure $p$ in the last 
shape equation in  \eqref{eqn:shape-equations}, 
where $\Delta \rho  =\Delta \rho_\mathrm{AC}=
\rho_\mathrm{A} -\rho_\mathrm{C}$ is the 
density difference between the capsule interior C  and the 
liquid phase A  for $s_0<\ell$ (where the capsule is in contact with 
liquid A) and  $\Delta \rho =\Delta
\rho_\mathrm{BC} = \rho_\mathrm{B} -\rho_\mathrm{C}$ 
the density difference between 
liquid phase B and capsule interior C for $s_0>\ell$ 
(where the capsule is in contact with 
liquid B).
Second, the matching conditions at the AB-interface have to be  modified 
if the angle $\vartheta$ is non-zero.
Then the force density  $\vec{f}_\sigma = \sigma( \sin\vartheta \vec{e}_z + 
   \cos\vartheta \vec{e}_r)$  acquires 
 a non-zero $z$-component $\sigma \sin\vartheta$, which enters
the Neumann triangle condition in $z$-direction, and 
the modified $r$-component $\sigma \cos\vartheta$ enters the 
Neumann triangle condition in $r$-direction. 

Thus,  our model can be easily extended to include all gravitational 
effects properly if larger, millimeter-sized 
capsules are considered that are no longer small 
as compared to the capillary length $L_c$
or the  gravitational  elastocapillary length $L_g$.

\section{Results}

In the remainder of this paper, we present 
and discuss numerical, quantitative results
 from solving the shape equations for different 
control parameters: the dimensionless bending 
modulus $\tilde{E}_\mathrm{B}$ or  F\"oppl-von-K\'{a}rm\'{a}n number
controlling the relevance of bending energy,
 the dimensionless surface tension $\tilde{\gamma} \equiv \gamma/Y_\mathrm{2D}$ 
controlling  the capsule elasticity from liquid droplet to 
elastic capsule,
and the dimensionless  interface load 
$\tilde{\sigma}\equiv \sigma/Y_\mathrm{2D} = L_{\sigma}/H$
or $\tilde{\sigma}/\tilde{\gamma} = \sigma/\gamma$
characterizing the strength of the deforming force from the 
AB-interface.
We will  neglect
gravitational effects, i.e., consider the typical situation 
that microcapsules are smaller than both gravitational capillary lengths,
$R_0 \ll L_g$ and $R_0 \ll L_c$.

 \begin{figure}[thb!]
 \centering
 \includegraphics[width=0.9\linewidth]{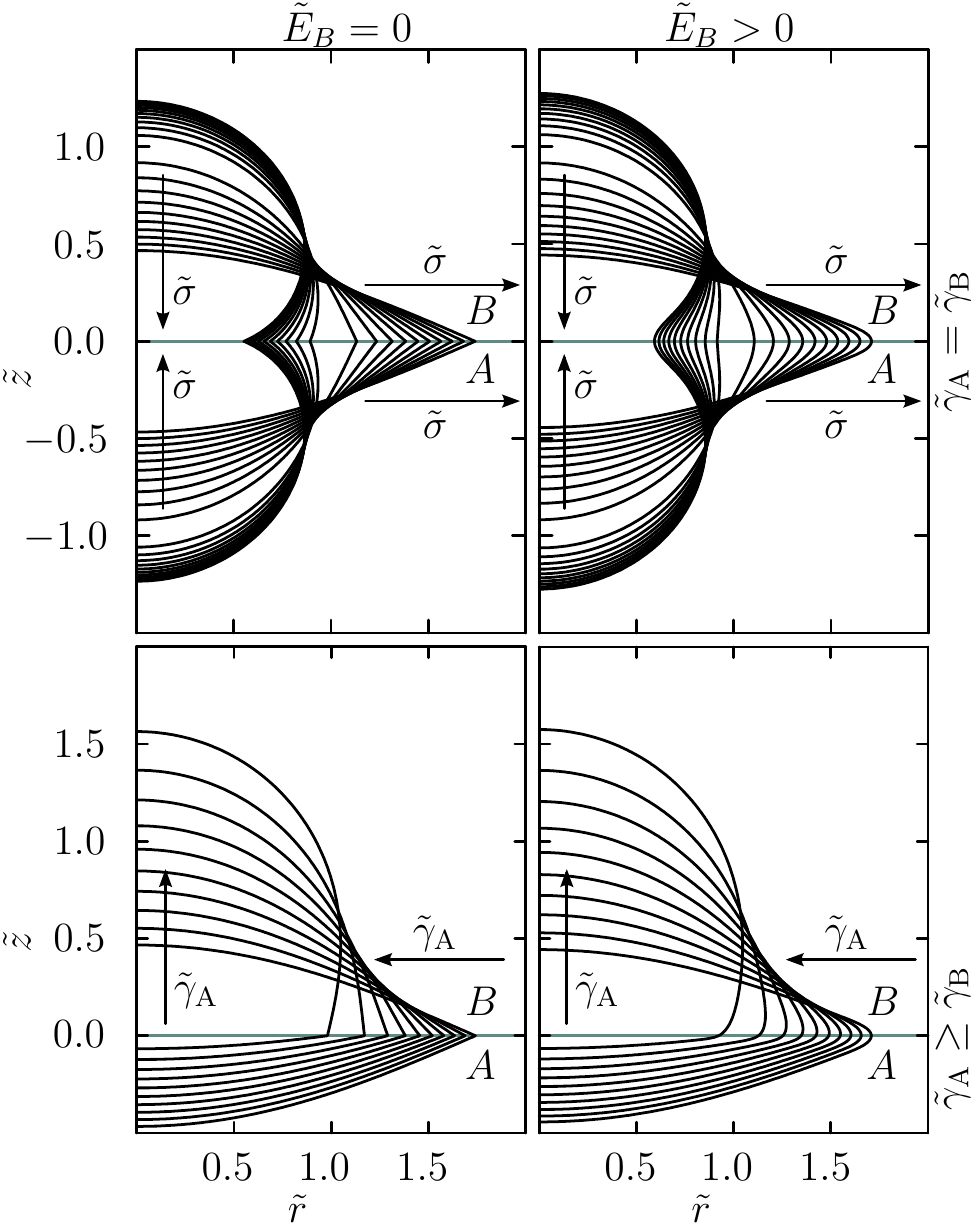}
 \caption{
  Shape evolutions for the Hookean membrane
   ($\tilde{E}_\mathrm{B}=E_B/Y_\mathrm{2D}R_0^2=0$, left)
  and the shell ($\tilde{E}_\mathrm{B}=10^{-3}$, right) both for 
  the symmetric case (upper row) and the asymmetric case (lower row).
  In the upper row we increase the interface 
 load from $\tilde{\sigma}=\sigma / Y_\mathrm{2D}=-1$
  (contractile, dumbbell-shaped) to $\tilde{\sigma}=1$ 
  (extensile, discus-shaped) at a constant ratio
  $\gamma_\mathrm{A}/\gamma_\mathrm{B} = 1$ (symmetric case).
  In the lower row we increase the ratio of surface
   tensions $\gamma_\mathrm{A}$ at the A-C interface
  (lower half-space) and $\gamma_\mathrm{B}$ at the B-C interface (upper 
half-space) 
  from $\gamma_\mathrm{A}/\gamma_\mathrm{B} = 1$ to
  $\gamma_\mathrm{A}/\gamma_\mathrm{B} = 10$ 
  at a constant interface load $\tilde{\sigma} = 1$.
  In contrast to the Hookean shell (right), 
  that shows a smooth and rounded edge at the interface,
  the Hookean membrane (left) exhibits a 
  sharp kink at the interface due to a vanishing shell thickness.
  Apart from this feature, shapes look rather 
   similar at small values of the dimensionless
  bending modulus $\tilde{E}_B$.
}
 \label{fig:num-solA}
\end{figure}

We also discuss analytical results for characteristics of the 
lens-like  droplet or capsule shapes,
where it is possible. 
We will show that 
the height or contact angle of the  lens
 is directly related to
the Young's modulus of the capsule shell (or the surface tension 
of a droplet), i.e., the dimensionless parameters 
$\sigma/\gamma$ characterizing the tensile force exerted
by the AB-interface and the dimensionless surface tension
 $\gamma/Y_\mathrm{2D}$ governing the crossover from a liquid 
droplet to an elastic Hooke membrane or shell.
The
maximal curvature at the ``tips'' of the rounded lens,
allows us to infer information about the bending modulus of the shell.

Finally, we can quantify the  adsorption 
energy gain by deformation relative to the adsorption energy gain of 
a hard undeformable particle, which is an important quantity for
applications of capsules as surface active agents, for example, 
in emulsification.

In Fig.\ \ref{fig:num-solA}, 
we  show numerically calculated 
evolutions of capsule shapes for varying 
interface loads  $\sigma$,
both for tensile  ($\sigma >0$) and 
contractile ($\sigma <0$)  tensions.
We show these evolutions 
both for symmetric ($\gamma_\mathrm{A} = \gamma_\mathrm{B}$) 
and asymmetric surface tension between capsule and phases A and B
and both for a Hookean membrane without bending rigidity and an 
elastic shell ($\tilde{E}_B=10^{-3}$). 
We clearly see the typical lens- or discus-like shapes for tensile 
interface loads. For contractile loads, a stable shape can only 
be obtained in the symmetric case, where we observe characteristic 
dumbbell shapes.

In the following 
paragraphs, we discuss aspects of the lens-like 
 shapes under tensile interface loads $\sigma$ quantitatively.
We focus on symmetric 
surface tensions $\gamma_\mathrm{A} = \gamma_\mathrm{B}=\gamma$
(shapes in the upper part of Fig.\ \ref{fig:num-solA})
and vary the dimensionless surface tension 
$\gamma/Y_\mathrm{2D}$ and bending modulus $\tilde{E}_\mathrm{B}$ to 
cover all  different surface
elasticities: the well-known liquid lenses for droplets
partially wetting the AB-interface ($\gamma/Y_\mathrm{2D} \to \infty$, 
$\tilde{E}_\mathrm{B}=0$),
elastic lenses for 
Hookean membranes  ($\gamma/Y_\mathrm{2D}$ finite, $\tilde{E}_\mathrm{B}=0$),
and rounded elastic lenses for microcapsules with 
Hookean shell elasticity
  ($\gamma/Y_\mathrm{2D}$ finite, $\tilde{E}_\mathrm{B}>0$).

 \begin{figure*}[t]
 \centering
  \includegraphics[width=\linewidth]{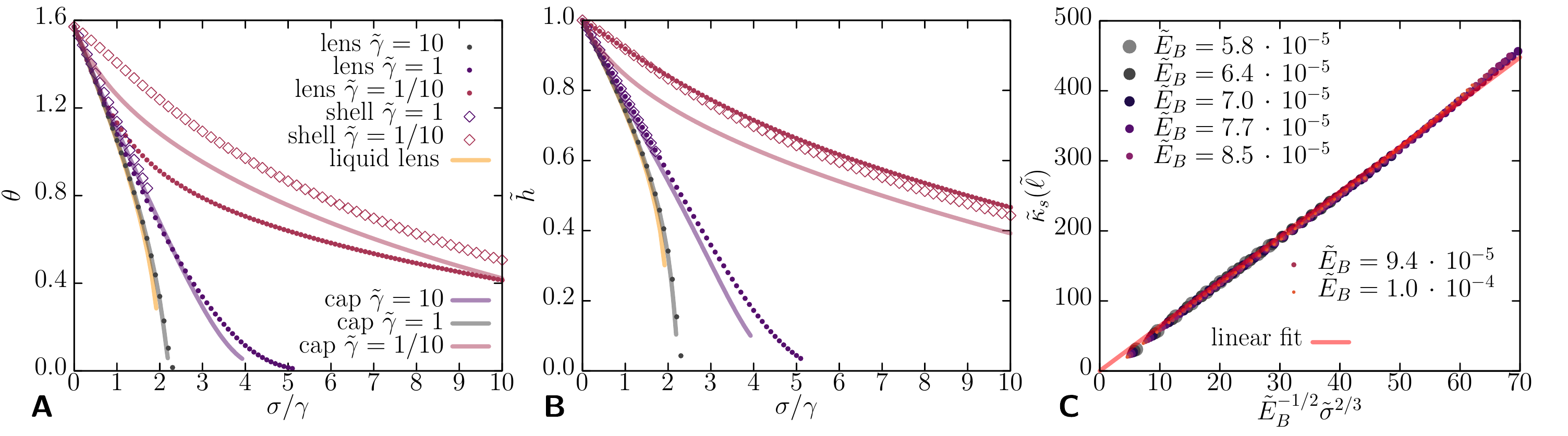}
 \caption
   {(A)
   Contact angle $\theta = \psi^-(\ell) = \pi-\psi^+(\ell)$ as a function of 
   the dimensionless tension $\tilde{\sigma}/\tilde{\gamma} = \sigma / \gamma$ 
   for the symmetric case ($\gamma_\mathrm{A} = \gamma_\mathrm{B}$):
   comparison of liquid lens result 
   \protect\eqref{eqn:wetting-angle} (yellow line) and spherical 
   cap approximation 
   \protect\eqref{eqn:cap-theory} (red, purple, and gray lines) 
   for the elastic lens 
   with numerical results from solving the shape equations 
  (red, purple, and gray dots).  
   Spherical cap approximation and numerical 
   results for the elastic lens are shown for three different regimes:
   (i) fluid regime
   $\tilde{\gamma} = \gamma /Y_\mathrm{2D}  \gg 1$ (gray), 
  (ii) crossover regime $\tilde{\gamma}
   \sim 1$ (purple), and (iii) elastic regime $\tilde{\gamma} \ll 1$ (red).
  In addition, we compare with   numerical results for shells 
   with $\tilde{E}_B = 10^{-3}$ (open quads) for the 
   crossover regime (purple), and the 
    elastic regime $\tilde{\gamma} \ll 1$ (red).
   For shells, we use an effective opening angle $\theta_{\rm eff}$, 
  which is obtained from the reduced height by the geometric 
  relation
   $\cos \theta_{\rm eff} = 2({1-\tilde{h}^3})/({2+\tilde{h}^3})$, 
   for spherical caps. 
     (B) Analogous comparison for the reduced height 
   $\tilde{h}=h/R_0$ (using the same symbols).
     Numerical results for shells (open quads) in 
  (A) and (B) show that even for shells of finite (though small) 
  thickness the spherical cap approximation 
   \protect\eqref{eqn:cap-theory} works well.
   (C) Numerical verification of the 
   scaling law 
   $\tilde{\kappa}_s \sim \tilde{E}_\mathrm{B}^{-1/2}\tilde{\sigma}^{2/3}$ 
   for elastic shells in the purely elastic
   regime $\tilde{\gamma} = 0$, see  eq.\ (\ref{eq:kappas}). 
   All quantities are given in dimensionless 
   form, $\tilde{\kappa}_s = \kappa_s R_0$, 
   $\tilde{\sigma} = \sigma/Y_\mathrm{2D}$, 
  and $\tilde{E}_B = E_B/Y_\mathrm{2D}R_0^2$.}
 \label{fig:wetting_exponent}
 \end{figure*}

\subsection{Droplets and membranes: lenticular shapes}

In the droplet regime (vanishing elastic moduli $Y_\mathrm{2D}$ and 
$E_\mathrm{B}$), the
correct shape is found from the Laplace-Young equation, 
according to which  the shape is assembled from 
spherical caps (shapes of constant mean curvature). 
For spherical caps 
the contact angle (between the tangent 
to the surface and the horizontal liquid-liquid interface) is related to 
their radius $R$ and  height $h$ by
\begin{align}
\cos \theta(\tilde{h}) = 1-\frac{\tilde{h}}{\tilde{R}(\tilde{h})}
 = 2\frac{1-\tilde{h}^3}{2+\tilde{h}^3}\,\text{.}
\label{eqn:wetting-angle}
\end{align}
Dimensionless lengths denoted with a tilde are measured 
in units of $R_0$.
The dimensionless  height $\tilde h$ of the spherical cap  
is its  maximal distance from the circular base of the cap.
For incompressible liquids as investigated in this paper
 the radius $\tilde{R}$ is related to the height $\tilde{h}$
of the spherical cap  by the 
volume constraint (see eq.\ \eqref{eqn:cap-volume-constraint_app}
in the Appendix), which gives the last equality 
in eq.\ (\ref{eqn:wetting-angle}). 
Balancing tensions at the three phase contact point, 
according to the Neumann triangle condition  \eqref{eqn:Young}
for force balance 
in $r$-direction, $\sigma = 2\gamma\cos\theta$,
we finally find that the height $\tilde{h}$ is given by
\begin{align}
\tilde{h} &= \frac{2^{1/3}(2-\sigma/\gamma)^{1/3}}{(4+\sigma/\gamma)^{1/3}}
\label{eqn:wetting-height}
\end{align}
for the liquid lens (see also eq.\ (\ref{eqn:h_app}) and its
derivation in the Appendix).
 For increasing the dimensionless  interface load $\sigma/\gamma$,
the height of the liquid lens decreases until we reach the transition 
to complete wetting for $\sigma=2\gamma$, where $\tilde{h}=0$ and 
the liquid lens becomes an (infinitely thin) wetting film.

For elastic membranes  with 
vanishing $E_\mathrm{B}$, but finite $Y_\mathrm{2D}$,
the presence of additional elastic stresses makes the Laplace-Young 
equation inapplicable. However, it is intuitively clear that adding 
a thin elastic membrane onto a droplet should not affect its shape 
drastically, 
such that the elastic lens should consist of approximate 
spherical caps as well. 
This is corroborated by our numerically calculated shapes in Fig.\ 
\ref{fig:num-solA} (left), which appear to be 
similar to  liquid lens shapes composed of two spherical caps. 
We will exploit this similarity for an analytical approximation. 

A closer inspection (see Appendix \ref{sec:elasticlens_app}) 
reveals that 
strictly spherical caps with hemispherical rest shape do 
not allow for local force balance over
the complete surface. However, we find an approximation analogous
to the droplet by assuming uniform (but anisotropic) 
strains along the contour. 
Geometrically, these strains are found to be
\begin{align}
\lambda_s &= \frac{2\theta(\tilde{h})}{\pi} \,\tilde{R}(\tilde{h})
   \hspace*{5mm} \text{and} \hspace*{5mm}
 \lambda_\phi = \tilde{R}_B = \tilde{R}(\tilde{h})\sin(\theta(\tilde{h})).
\label{eqn:spherical-strains}
\end{align}
Inserting these approximative strains in the constitutive relations 
\eqref{eqn:constitutive-laws} we obtain 
the stress $\tau_s$  as a function of $\tilde{h}$.
Using the Neumann 
condition \eqref{eqn:Young} for force balance 
in $r$-direction, 
 we then find (see Appendix \ref{sec:elasticlens_app})
\begin{align}
\sigma &= 2\tau_s(\tilde{h})\cos\theta(\tilde{h})
= 4\tau_s(\tilde{h}) \frac{1-\tilde{h}^3}{2+\tilde{h}^3} ,
\label{eqn:cap-theory}
\end{align}
where $\cos\theta(\tilde{h})$ is given by the geometric relation 
\eqref{eqn:wetting-angle}.
Similarly to \eqref{eqn:wetting-height}, solving this equation for $\tilde{h}$
 gives the reduced height $\tilde{h}$ and, thus, 
also the contact angle $\theta$ 
as a function of $\sigma/\gamma$ and $\gamma/Y_\mathrm{2D}$.

We compare our  analytical results
for height $\tilde{h}$ and  opening angle $\theta$
 to numerical solutions of the 
 shape equations \eqref{eqn:shape-equations}
 in Fig.\ \ref{fig:wetting_exponent}(A,B).
 The result \eqref{eqn:wetting-height} for the height of liquid lenses 
is in good agreement with solutions for 
elastic membrane lenses in the corresponding
limit $\gamma \gg Y_\mathrm{2D}$. 
Solutions of \eqref{eqn:cap-theory} for the height 
are in good agreement 
with numerical simulations within the range of the fluid
and the crossover regime, i.e., for $\gamma \geq Y_\mathrm{2D}$. 
Even for $\gamma \leq Y_\mathrm{2D}$
we find acceptable agreement.
Note that only the dimensionless height $\tilde{h}$ can directly
be determined for elastic shells, since the shape is rounded 
at the AB-interface.
The solutions we obtain from \eqref{eqn:cap-theory} violate, however, 
the force balance condition, which explicitly 
demonstrates that already a simple 
Hookean stretching energy leads to non-trivial shapes.

\subsection{Elastic shells: interface curvature}

Microcapsules with elastic shells, i.e., 
in the presence of an additional  bending rigidity,
exhibit lens-like  shapes with rounded kinks at the AB-interface.

For decreasing  bending moduli the rounded kink at the AB-interface 
becomes increasingly sharp and approaches the elastic lens shape
suggesting a systematic  relation between the interface curvature 
and the bending modulus  which 
might allow for inferring the bending modulus from the shape profile.
We quantify this in Fig.\ \ref{fig:wetting_exponent}(C)
showing the curvature $\kappa_s$ at the liquid-liquid interface
as a function of $E_\mathrm{B}^{-1/2} \sigma^{3/2}$,
where we clearly find a linear relationship
implying $\kappa_s \propto E_\mathrm{B}^{-1/2} \sigma^{3/2}$ in the 
elastic regime $\gamma/Y_\mathrm{2D} \approx 0$. 
This scaling can be rationalized by an argument, which 
is similar to the  Pogorelov  theory for the rounding of 
the rim of a buckled elastic spherical shell be bending 
rigidity \cite{pogorelov1988bendings,LL7}.
Details of this argument  can be found in Appendix 
\ref{sec:Pogorelov_app}. 
The essential idea is to consider the rounding of the sharp cusp present 
for an elastic lens with contact or opening angle $\theta$ and 
 a complementary angle
$\alpha=\pi/2-\theta$,
 after introduction of a finite bending rigidity.
 Thus, we only have 
to consider the deviation from the rather simple elastic lens shape on a 
typical length scale $\xi$ that is found from balancing stretching and bending 
contributions, from which we find the indicated scaling 
of the curvature $\kappa_s = \alpha/\xi$.
The change in the bending energy scales as 
$U_B = E_\mathrm{B}R_D\alpha^2/\xi^2$ whereas
the interfacial contribution scales as 
$U_\sigma = \pi R_D\sigma \xi \alpha$, where 
$R_D$ is the radius of the interface cross-section.
Balancing both energies leads to 
$\xi \sim \alpha^{1/2}{E_\mathrm{B}}^{1/2}{\sigma}^{-1/2}$ or 
$\kappa_s\sim \alpha^{1/2} E_\mathrm{B}^{-1/2} \sigma^{1/2}$, 
which already proves $\kappa_s \propto  E_\mathrm{B}^{-1/2}$, but still depends 
on the angle $\alpha$. 
The scaling of $\alpha$ 
is obtained from the Neumann condition \eqref{eqn:cap-theory},
$\sigma = 2\tau_s(\tilde{h})\sin\alpha$,
and  depends on whether the surface
tensions (fluid regime, $\alpha \sim \sigma/\gamma$)
 or the Young's modulus (elastic
regime, $\alpha \sim (\sigma/Y_\mathrm{2D})^{1/3}$) 
constitute the dominant stretching
force.
This results in the final result for the 
curvature scaling, 
\begin{align}
\kappa_s &\sim 
\begin{cases} E_\mathrm{B}^{-1/2}\sigma^{2/3}\,Y_\mathrm{2D}^{-1/6}, &
         \gamma \ll Y_\mathrm{2D}
         \hspace*{5mm} \text{(elastic)} \\
                   E_\mathrm{B}^{-1/2}\sigma \gamma^{-1/2}, & 
       \gamma \gg Y_\mathrm{2D} \hspace*{5mm} \text{(fluid)},
\end{cases}
\label{eq:kappas}
\end{align}
which is in full agreement with the numerical 
results in Fig.\ \ref{fig:wetting_exponent}(C). 
Note that the dependence on  the bending modulus,
 $\kappa_s \propto  E_\mathrm{B}^{-1/2}$ is universal, i.e.,
independent of whether we are in the fluid or elastic  regime. 
Switching from $E_\mathrm{B}=0$ to a finite $E_\mathrm{B}$ 
leads to a finite  $\alpha$
and, thus, rounded edges.

Figs.\ \ref{fig:wetting_exponent}(A) and (B) show that 
the analytical result (\ref{eqn:cap-theory}) for the 
 capsule height $h$ and the 
effective opening angle, which is obtained via the geometrical 
relation 
$\cos \theta_{\rm eff} = 2({1-\tilde{h}^3})/({2+\tilde{h}^3})$ for 
spherical caps, remain a good approximation 
also for microcapsules with elastic shells for small 
dimensionless bending rigidities 
($\tilde{E}_\mathrm{B} \sim 10^{-3}$ in Figs.\ \ref{fig:wetting_exponent}(A)
and (B)). 
This is also  evident from 
   Fig.\ \protect\ref{fig:num-solA}, where 
   shell and lens heights differ only slightly.
Therefore, the parameters $\sigma/\gamma$ and $\gamma/Y_\mathrm{2D}$
can  be inferred from measurements of capsule heights 
over a wide range of elastic parameters by fitting with 
 eq.\ (\ref{eqn:cap-theory}) for the measured capsules heights.

Our result (\ref{eq:kappas})
 for the curvature radius $1/\kappa_s$ at the rounded 
tip  of an elastic shell  then also 
suggests that the classical Neumann condition
for symmetric liquid droplets, 
$\sigma = 2\gamma \cos \theta$, still holds for elastic shells and lenses
 as long as the curvature radius $1/\kappa_s$ at the rounded 
tip is small compared to the radius $R_0$ of the capsule, i.e., 
for $1/\kappa_sR_0 \sim \tilde{E}_\mathrm{B}^{1/2}\tilde{\sigma}^{-2/3} \ll 1$ 
(in the elastic regime  $\tilde{\gamma} < 1$), 
 if the approximative uniform 
stress $\tau_s(\tilde{h})$ is used as capsule-liquid  surface tension $\gamma$,
 and if the  effective Neumann
angle $\theta_{\rm eff}$ is used for $\theta$. 
Similar observations have been made for the shapes of adhered  vesicles
\cite{Seifert1990},
which are also governed by the Young equation for liquid droplets 
with en effective surface 
tension as long as the contact curvature radius is small compared 
to the vesicle size. 
In Fig.\ 
\ref{fig:wetting_exponent}(B), where $\tilde{E}_B=10^{-3}$ (quads) and 
where we focus on the elastic regime 
$\tilde{\gamma} = 1/10$ and $\tilde{\gamma} = 1$, the 
 condition 
$1/\kappa_sR_0 \sim \tilde{E}_\mathrm{B}^{1/2}\tilde{\sigma}^{-2/3} \ll 1$ 
of small curvature radii 
is already fulfilled for 
$\tilde{\sigma}/\tilde{\gamma} \gg 10^{-5/4}$ and 
$\tilde{\sigma}/\tilde{\gamma} \gg 10^{-9/4}$, respectively.

\subsection{Adsorption energy  enhancement}

Soft particles at liquid-liquid interfaces are efficient  
emulsifiers because they stretch during adsorption~\cite{style2015adsorption}.
During deformation at the liquid-liquid interface a soft 
particle assumes a lens-like shape~\cite{mehrabian2016soft,geisel2015hollow},
which increases the occupied interface area and, thus, decreases
the interfacial energy, while the elastic energy increases.
Typically, the elastic energy cost is smaller than the 
energy gain due to spreading within the interface, meaning
that spreading is energetically preferable. 
The higher the interfacial energy gain $-\sigma\pi r^2(\ell)$
compared to the elastic cost (integrated energy density $w(s_0)-\gamma$),
the more stable the interface becomes.
The sum of the above two contributions is what we refer 
to as the {\it adsorption energy}.
In Ref.\ \citenum{bresme2007nanoparticles} the adsorption
 stability of nanoparticles
at liquid-liquid interfaces has been investigated as a function of the particle
shape, where it turns out that oblate shapes are most stable due to the high
area occupation within the interface.
Therefore, hollow elastic capsules with a thin elastic shell, 
which are much softer than filled particles, are very attractive 
candidates to improve emulsification further.  

Using our numerical results we can quantify the increase in adsorption 
energy as a function of the softness of the capsule. 
The adsorption energy of a deformable 
 capsule is given by the total energy gain 
if the capsule is  moved from one of the  liquid phase 
to the liquid-liquid interface,
\begin{equation}
\Delta E_\mathrm{soft} = -\sigma \pi r^2(\ell)  
+\int (w(s_0)-\gamma) \,\mathrm{d}A_0,
\label{eqn:Esoft}
\end{equation}
where $\pi r^2(\ell)$ is 
the occupied circular cross-section area within the liquid-liquid
interface plane and the last term is the 
elastic energy including stretching energy, bending energy, and 
the change in surface energy $\gamma (A-A_0)$.
Each capsule moving to  the liquid-liquid interface lowers the 
interfacial energy by $\Delta E_\mathrm{soft}$ and, 
thus, decreases the effective 
surface tension of the liquid-liquid interface. 

\begin{figure}[t]
\includegraphics[width=\linewidth]{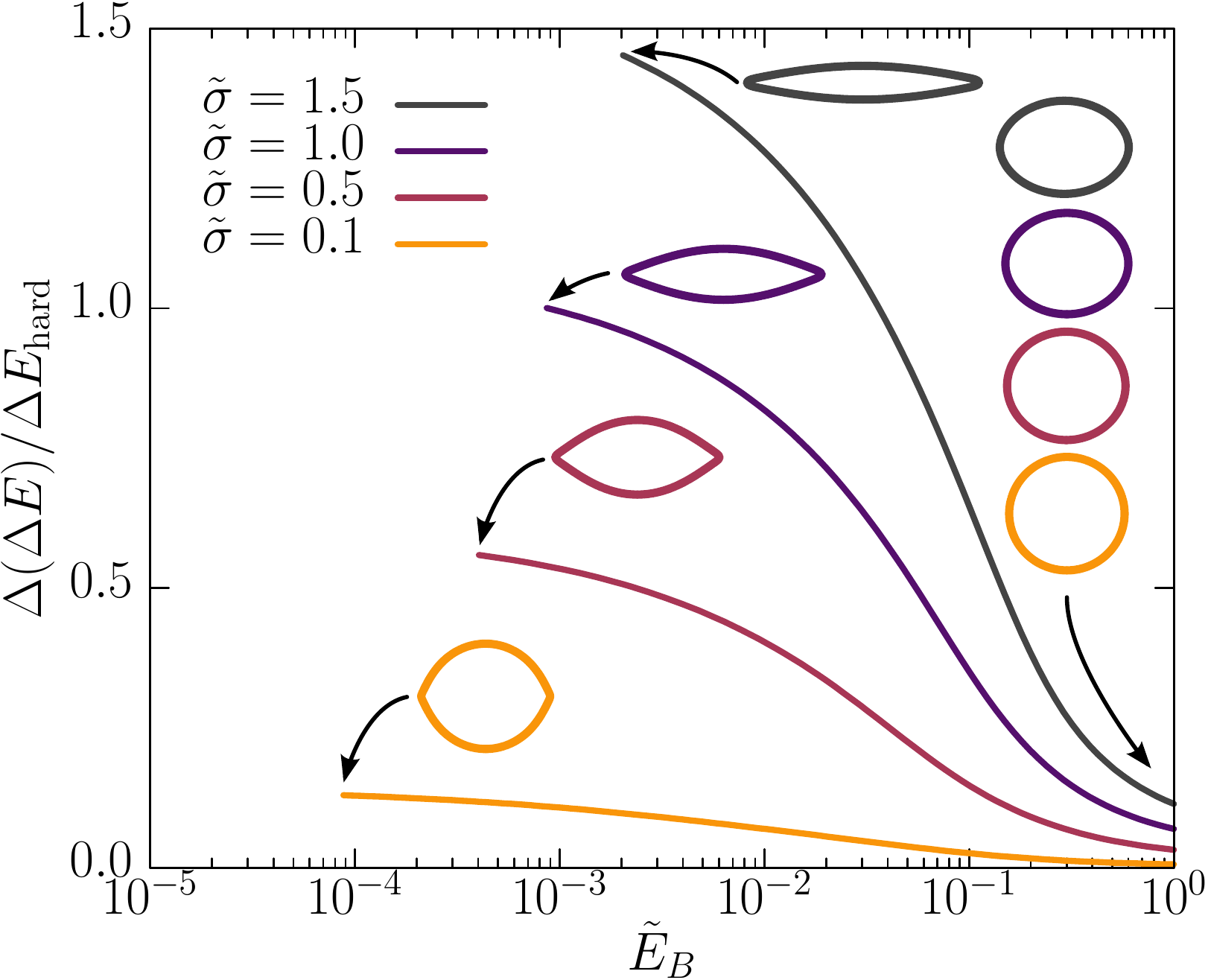}
\caption{
  Normalized adsorption energy difference 
  $\Delta (\Delta E)/ \Delta E_\mathrm{hard}$ between soft and hard particles 
  as a function of the 
  dimensionless bending rigidity $\tilde{E}_\mathrm{B} \sim H^2/R_0^2$ 
  for different values of the dimensionless liquid tension 
  $\tilde{\sigma} = \sigma/Y_\mathrm{2D}$. 
  $\Delta (\Delta E)/ \Delta E_\mathrm{hard} = 1$ is equivalent
  to a $100\,\%$ enhancement of the adsorption energy difference
  of a soft capsule compared to a hard particle.
  The adsorption energy is strongly enhanced for hollow capsules
  with small thickness to radius ratios $H/R_0$ and for large 
  liquid tensions. 
  For the above measurement we used $\gamma/Y_\mathrm{2D} = 1/10$.
  The insets show exemplary shapes for different values of 
  $\tilde{E}_B$ and $\tilde{\sigma}$.
}
\label{fig:adsorption}
\end{figure}

This effect is stronger for hollow soft elastic capsules as compared to
hard particles and becomes more pronounced with decreasing 
thickness of the capsule.  
For hard spherical particles of equal size 
we have an adsorption energy  
$\Delta E_\mathrm{hard} = - \sigma\pi R_0^2$.
In Fig.\ \ref{fig:adsorption} we numerically quantify
the adsorption energy {\it difference} between soft and hard particles, 
\begin{align}
\Delta (\Delta E)\equiv \Delta E_\mathrm{soft}-\Delta E_\mathrm{hard},
\end{align}
as a function of the dimensionless 
bending modulus $\tilde{E}_\mathrm{B} = H^2/12(1-\nu_\mathrm{2D}^2)R_0^2$,
which characterizes the thickness of the capsule, 
see eq.\ (\ref{eqn:EB}). 
This allows us to quantify  the relative 
enhancement of the adsorption 
energy due to capsule softness by decreasing $\tilde{E}_\mathrm{B}$
or the capsule thickness. 
For $\tilde{E}_\mathrm{B}\sim 10$ 
(corresponding to $H\sim R_0$ and $\nu_\mathrm{2D}=1/2$) 
the results 
should become similar to the energy gain for soft filled particles
as they have been considered in Refs.\ 
 \citenum{mehrabian2016soft,geisel2015hollow,style2015adsorption}.
Fig.\  \ref{fig:adsorption} then clearly shows that hollow 
capsules are much more efficient as emulsifiers than
 filled soft particles (or even a filled hard particle) of 
equal size.

\section{Discussion and conclusion}

We investigated shapes of deformed microcapsules adsorbed at
liquid-liquid interfaces. We mainly focused on the situation where 
capsules are stretched by a liquid-liquid interface outside the 
capsule, see Fig.\ \ref{fig:intro}(A) but our numerical methods 
(including matching conditions) are equally applicable to 
compressive tension, which arise, if a liquid-liquid interface 
is formed inside the capsule, see Fig.\ \ref{fig:intro}(B).
We demonstrated this in Fig.\ \ref{fig:num-solA}, where we 
present numerical results for  both cases.

Gravitational effects can be generically 
neglected for  micrometer-sized capsules because both the 
capillary length $L_c = (\sigma/\Delta \rho g)^{1/2}$ and 
the  gravitational  elastocapillary length
 $L_g = (Y_\mathrm{2D}/\Delta \rho g)^{1/2}$ are in the millimeter range
exceeding capsule size. We showed in section \ref{sec:grav},
how  our model can be extended to include all gravitational 
effects if larger, millimeter-sized 
capsules are considered in future work.

The deformability of hollow microcapsules by the liquid-liquid
interfacial tension mainly depends on their thickness $H$.
For hollow capsules with a soft shell
strains become  large if $\sigma > Y_\mathrm{2D} \sim 
   Y_\mathrm{3D}H$, i.e., if 
shell thickness $H$ is sufficiently small compared 
to the elastocapillary length $L_{\sigma} = \sigma/Y_\mathrm{3D}$.
For a typical soft capsule  shell material with 
$Y_\mathrm{2D}\sim 10^{-2}...10^{-1} \,\mathrm{N/m}$, liquid-liquid 
interfacial tensions $\sigma >   10^{-2} \,\mathrm{N/m}$ are 
sufficient to induce considerable (but not large) 
strains, which can be realized
by oil-water interfaces. 

Shapes of deformed capsules with spherical rest shapes are
lens-like if they are stretched at a liquid-liquid interface;
in the presence of bending rigidity the edge of the 
lenticular shape is rounded. 
Neglecting gravitation
we derived shape equations and matching conditions at the liquid-liquid 
interface for the numerical calculation of these shapes
for extensible shells of finite thickness and  constant volume, 
as well as for two important limiting cases, 
namely elastic lenses (zero bending modulus) and liquid lenses 
(zero Young's and bending modulus, only surface tension). 
We calculated numerical solutions for each of these cases,
see Fig.\ \ref{fig:num-solA}.
We also 
 derived  analytical approximations 
for characteristics of the deformed capsule shapes in agreement with 
our numerical results.

Our approximative theory  based on  spherical cap shapes can be used 
to determine Young's modulus $Y_\mathrm{2D}$ and 
the pressure $p_0$ from a single 
measurement of the cap height $h$ or the contact angle $\theta$, 
if the surface tensions $\sigma$ and $\gamma$ are known, see
Fig.\ \ref{fig:wetting_exponent}(A,B). 
Two height measurements of the same capsule at different 
surface tensions $\sigma$ or at different capsule volumes $V_0$ 
could be used to determine both Young's modulus $Y_\mathrm{2D}$ 
and the capsule-liquid surface tension $\gamma$
(assuming values for the Poisson ratio, for example, $\nu_\mathrm{2D}=1/2$). 

This could be further extended, and 
our scheme for numerical calculation of shapes  can, in
principle, be used for elastometry, i.e.,  
to determine  elastic moduli from an experimentally
acquired image by fitting numerical solutions of the shape equations to a set
of contour points extracted from the image.
Other elastometry methods following the same philosophy are 
the study of deformations of pendant capsules under volume changes 
to obtain elastic moduli
as investigated in Refs.\ \citenum{Knoche2013,Nagel2017,hege2017elastometry},
the study of the edge curvature of a buckled shapes to obtain 
the bending modulus  \cite{Knoche2011}, or the study of shapes 
of osmotically buckled capsules to infer the osmotic 
pressure \cite{Knoche2014osmotic}.

For shells of finite
thickness we also found an analytical result for the maximal curvature 
at the ``tips'' of the rounded lens  in terms of the 
bending modulus, Young's modulus and the interface load, see 
eq.\ (\ref{eq:kappas}) and Fig.\ 
\ref{fig:wetting_exponent}(B).
Also this analytical  result 
can be  used in several ways to 
extract information 
on elastic moduli from a measurement of the 
shell's curvature $\kappa_s$ at the interface and if the interface load
$\sigma$ from the liquid-liquid surface tension is known.
If  Young's modulus $Y_\mathrm{2D}$ of the capsule material is known 
(for example, from other elastometry methods 
\cite{Knoche2013,Nagel2017,hege2017elastometry})  we can determine 
the bending modulus by 
\begin{align}
E_\mathrm{B} &= 40.9 \, 
  Y_\mathrm{2D}^{-1/3} \sigma^{4/3}\kappa_s^{-2}.
\label{eqn:EBkappa1}
\end{align}
If the shell thickness $H$ is known 
 we can determine 
 bending modulus or Young's modulus by 
\begin{align}
\begin{split}
E_\mathrm{B} &= 9.3 \, H^{1/2}\sigma \kappa_s^{-3/2},\\
Y_\mathrm{2D} &= 84.1 \, H^{-3/2}\sigma \kappa_s^{-3/2},
\end{split}
\label{eqn:EBkappa}
\end{align}
where we used $\nu_\mathrm{2D} = 1/2$.
The numerical prefactors in eqs.\  (\ref{eqn:EBkappa1}) and 
(\ref{eqn:EBkappa}) were
determined by a linear fit to the numerical results
and we used relation \eqref{eqn:eb}  for $E_\mathrm{B}$
and $Y_\mathrm{2D}$.
Note that $\sigma
\kappa_s^{-3/2} = \mathrm{const}$, which is why it makes sense to vary the
interface load $\sigma$ in order to improve the statistical significance of
such a measurement.  This could be achieved by, e.g., adding surfactants to one
of the liquid phases $\mathrm{A}$ or $\mathrm{B}$ that decrease $\sigma$ with
increasing surfactant concentration. The dependence of $\sigma$ and the
surfactant concentration can be determined, e.g., 
 in a pendant drop tensiometer.
The curvature $\kappa_s$ can be obtained from analyzing 
capsule images and  fitting a circle to 
the capsule edge at the liquid-liquid interface.
Using typical values  for soft microcapsules 
($E_\mathrm{B}=10^{-16}\dots 10^{-14}\,\mathrm{Nm}$,
 $Y_\mathrm{2D}=10^{-2}\dots 10^{0} \,\mathrm{N/m}$ corresponding to 
$H=0.03\dots 3\,{\rm \mu m}$) and 
 a  liquid-liquid 
interfacial tension (e.g.\ oil-water) of 
$\sigma \sim  5\cdot 10^{-2} \,\mathrm{N/m}$, we expect 
curvature radii
$1/\kappa_s \sim 0.005\dots 0.1\, {\rm \mu m}$ according to 
(\ref{eqn:EBkappa1}), which can be measured optically 
as has been demonstrated in Ref.\ \citenum{Jose2014}.

In conclusion, the results presented in this paper 
 allow us, in principle,  to determine the full
set of elastic constants from shape profiles of elastic capsules adsorbed to
liquid-liquid interfaces. 
To realize this setup experimentally, several problems have to be 
solved. 
One main problem will be to find 
a liquid-liquid interface  with  sufficiently high surface tension such that 
the condition   $\sigma > |\gamma_\mathrm{A} - \gamma_\mathrm{B}|$
for adsorption of the capsule to the interface is fulfilled and such that 
the observed strains $\Delta R/R_0 \sim \sigma/Y_\mathrm{2D}$ and, thus, 
the capsule deformation are observable. 
Another problem might be leakage through the capsule membrane, 
while we assume a fixed volume in the present calculation. 
A known shrinking volume can, however, be included in the present 
approach by introducing 
a corresponding decreasing internal pressure $p_0$.

Finally, we could show that hollow elastic microcapsules can be  much more 
effective in reducing the interfacial energy than filled soft particles
or even hard particles
of the same size, see Fig.\ \ref{fig:adsorption}. 
 During
capsule deformation at the liquid-liquid interface into a lens-like shape the
adsorption energy, which results from the 
balance of  the  occupied liquid-liquid interface area and the elastic energy 
of the capsules shell,
increases significantly.
In Fig.\  \ref{fig:adsorption} we compare hollow capsules, filled soft
particles and hard particles of equal size $R_0$. 
The effectiveness of hollow capsules in reducing the interfacial energy
can also be appreciated by comparing with particles of equal 
deformability:
A hollow capsule with shell thickness $H$  exhibits a similar 
elastic deformation as a filled soft particle of size $H$. 
The size of the hollow microcapsule can be made much larger, however, 
in order to achieve a much larger  occupied liquid-liquid interface area
as compared to the filled particle. 
This leads to much higher adsorption energies at comparable 
softness.
We conclude  that hollow microcapsules could be much more efficient 
in foam and  emulsion stabilization  than  filled 
particles of comparable size or comparable softness.

\section*{Conflict of interest}
There are no conflicts to declare.

\section*{Acknowledgements}
We acknowledge financial support by the Deutsche Forschungsgemeinschaft 
via SPP 1726 ``Microswimmers'' (KI 662/7-1).

\bibliography{lit}

\newpage
\appendix

\section{Shape equation approach}
\label{app:shape}

In this section, we give details on the derivation of 
 the shape equations (\ref{eqn:shape-equations}) by 
  variational energy minimization.
We give a detailed derivation of the matching conditions 
at the liquid-liquid interface and present details 
on the numerical solution of the shape equations.

\subsection{Shape equations}
\label{app:shape_equations}

\subsubsection{Variational approach}
\label{app:variational}

\begin{figure}
\centering
\includegraphics[width=\linewidth]{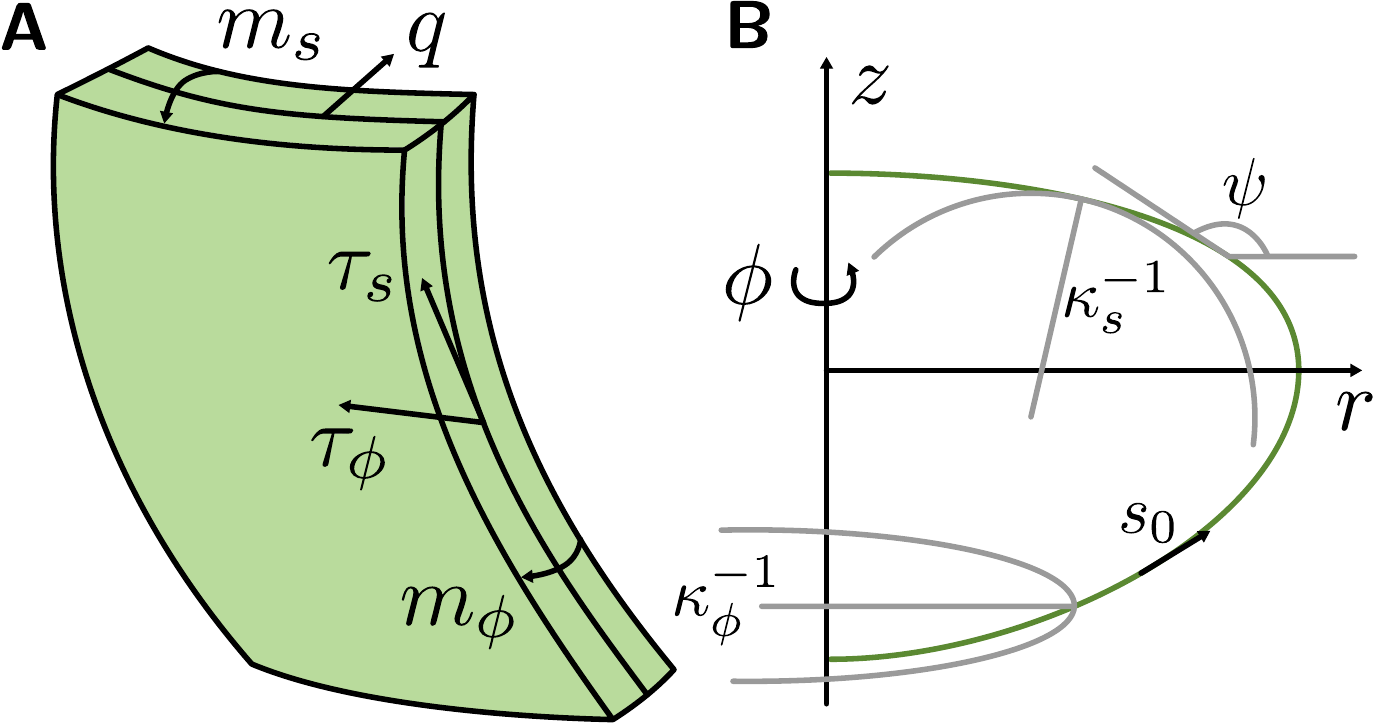}
\caption{
(A) Forces and torques acting on an
  infinitesimal membrane patch. Both forces and torques are integrated over
  the thickness $H$ of the material, which is assumed to be
  constant.
  (B) Shape profile given by the shape
  equations. The shape profile generates a surface of revolution describing a
  three-dimensional axisymmetric object. The curvatures $\kappa_\phi$ and
  $\kappa_s$ determine the local bending energy. 
}
\label{fig:interface-patch}
\end{figure}

We calculate the first variation and the resulting 
Euler-Lagrange equations for the free energy functional 
(\ref{eqn:Gvar})
by variation with respect to the 
 functions $r(s_0)$ and $\psi(s_0)$.
First we perform the variation of 
\begin{align}
G_0&= \int  w(s_0) \mathrm{d}A_0 - p_0V
   =   \int_0^{L_0} \mathrm{d}s_0  g ~~~~\mbox{with}
\nonumber\\
   g&\equiv  
      2\pi r_0 w
     -p_0 \pi r^2  \lambda_s  \sin\psi 
\label{eqn:G0var}
\end{align}
in the absence of the 
interface load $E_\sigma = -\sigma \pi r^2(\ell)$
(we repeat this calculation 
from Appendix A in Ref.\ \citenum{Knoche2011}
for completeness of the presentation), 
in order to derive the shape equations for the upper and 
lower part of the capsule. 
In Appendix \ref{app:match} we  include $E_\sigma$ to  
derive the matching conditions at the three phase contact line 
joining upper and lower part of the capsule.
Assuming an incompressible liquid within the capsule and an impermeable
membrane leads to conservation of the capsule volume
\begin{align}
V &= \int \pi r^2 \mathrm{d}z = 
   \int_0^{L_0}\pi r^2  \lambda_s  \sin\psi \mathrm{d}s_0 
   =V_0= \frac{4\pi}{3}R_0^3,
\label{eqn:vol}
\end{align}
where $R_0$ is the radius of the spherical equilibrium shape of 
the capsule. 
The volume constraint $V= V_0=\mathrm{const}$
is  included by adjusting the hydrostatic
pressure $p_0$ as a Lagrange parameter.

Using the definition of the stresses and bending moments 
(see first equalities in eqs.\ (\ref{eqn:constitutive-laws});
  in the following, $f'\equiv df/ds_0$) for the 
variation of the elastic energy density,
\begin{align}
\delta w &= \lambda_\phi \tau_s \delta e_s + \lambda_s \tau_\phi \delta e_\phi 
    + \lambda_\phi m_s \delta K_s + \lambda_s m_\phi \delta K_\phi
 ~~~\mbox{with}
\nonumber\\
   \delta e_s &= \frac{\delta r'}{\cos\psi} + \lambda_s \tan\psi \delta \psi
~,~~ 
  \delta e_\phi = \frac{1}{r_0}\delta r,
\nonumber\\
 \delta K_s &= \delta\psi'
~,~~ 
  \delta K_\phi = \frac{\cos\psi}{r_0}\delta\psi,
\label{eqn:deltaw_app}
\end{align}
and 
\begin{align}
   \delta( r^2\lambda_s \sin\psi) &= 
   2r \lambda_s \sin\psi \delta r + 
   r^2 \tan\psi \delta r' + 
    \frac{\lambda_s r^2}{\cos\psi}\delta \psi
\label{eqn:deltaV_app}
\end{align}
for the variation of the volume, 
we find the first variation of the free energy $G$
\begin{align*}
   \delta G_0 &= 
   \int_0^{L_0}  \mathrm{d}s_0 2\pi \Bigg( \delta r
   \Big\{
    \lambda_s \tau_\phi -  p_0 r\lambda_s \sin\psi 
  \Big\}
   \nonumber\\
  &  +\delta r'\Big\{
       \frac{r \tau_s}{\cos\psi}  -\frac{p_0}{2} r^2 \tan\psi 
    \Big\}
\nonumber\\
   & +  \delta \psi\Big\{ 
      r \tau_s \lambda_s \tan\phi  
   -\frac{p_0}{2} \frac{\lambda_s r^2}{\cos\psi} 
   +  \lambda_s m_\phi \cos\psi 
   \Big\}
  + \delta \psi' \Big\{
       r m_s 
       \Big\} \Bigg)
\end{align*}
or 
\begin{align}
 \delta G_0 &= \int_0^{L_0}  \mathrm{d}s_0  
  \left( \delta r g_r + \delta r' g_{r'} +
      \delta \psi g_\psi + \delta \psi' g_{\psi'} \right)~~\mbox{with}
\label{eqn:deltaG_app}\\
   g_r &\equiv 2\pi \lambda_s(\tau_\phi -  p_0 r \sin\psi)
~,~~ g_{r'} \equiv  2\pi (-rq\sin\psi\ + r\tau_s\cos\psi)
\nonumber\\
  g_\psi &\equiv  2\pi \lambda_s(-r  q +  m_\phi \cos\psi )
~,~~
  g_{\psi'} \equiv    2\pi r m_s,
\label{eqn:gdef}
\end{align}
where we also defined
\begin{equation}
    q \equiv -\tau_s \tan\psi + \frac{1}{2} p_0 \frac{r}{\cos\psi}.
\label{eqn:q0}
\end{equation}
Partial integration in eq.\ (\ref{eqn:deltaG_app}) 
gives two standard Euler-Lagrange equations 
\begin{align}
 {\rm (i)}~~  g_r &= \frac{\mathrm{d}}{\mathrm{d} s_0} g_{r'},
~~,~~
 {\rm (ii)}~~ g_\psi = \frac{\mathrm{d}}{\mathrm{d} s_0} g_{\psi'}
 \label{eqn:EL}
\end{align}
on the interval $s_0\in [0,L_0]$.
Using  $\lambda_s = \mathrm{d}s/\mathrm{d}s_0$ and the 
geometric relations
$\mathrm{d}r/\mathrm{d}s = \cos\psi$ and 
$\mathrm{d}\psi/\mathrm{d}s=\kappa_s$,
which follow from the definition of the slope angle $\psi$
(see Fig.~\ref{fig:interface-patch}(B)),
these Euler-Lagrange equations can be re-arranged to give 
\begin{align}
\begin{split}
 {\rm (i)}~~0 &= -\frac{\cos\psi}{r}\tau_\phi
      +\frac{1}{r}\frac{d(r\tau_s)}{ds}-\kappa_s q,\\
{\rm (ii)}~~0 &= \frac{\cos\psi}{r}m_\phi-\frac{1}{r}\frac{d(rm_s)}{ds}-q,
\end{split}
\label{eqn:equilibrium_var}
\end{align}
which represents (i) stress equilibrium in tangential direction 
and (ii) torque (bending moment) balance  \cite{Knoche2011,Libai1998}
that has to hold for any interface patch, 
 see Fig.~\ref{fig:interface-patch}(A).
Moreover, the definition (\ref{eqn:q0}) of the 
normal transverse shear force density $q$  can be used to derive (together
with (i)) the normal stress equilibrium  \cite{Knoche2011,Libai1998}
\begin{align}
 {\rm (iii)}~~0 &= -p_0 + \kappa_\phi\tau_\phi + \kappa_s\tau_s 
   + \frac{1}{r}\frac{d(rq)}{ds},
\label{eqn:equilibrium_var2}
\end{align}

In order to obtain a closed set of shape equations, we 
also use the geometric relations 
\begin{align}
\begin{split}
r'(s_0) &{=} \lambda_s \cos\psi, \hspace{2mm} 
 z'(s_0) {=} \lambda_s \sin\psi, \hspace{2mm} 
 \psi'(s_0) {=} \lambda_s \kappa_s,
\end{split}
\label{eqn:geometry}
\end{align}
(see Fig.~\ref{fig:interface-patch}(B)).  Using $\lambda_s =
\mathrm{d}s/\mathrm{d}s_0$ to relate deformed to undeformed arc length,
eqs.\ (\ref{eqn:equilibrium_var}), (\ref{eqn:equilibrium_var2}), and
(\ref{eqn:geometry}) are equivalent to the shape equations
(\ref{eqn:shape-equations}) in the main text.  
To convert these shape equation
system into actual geometric shapes constitutive relations are needed, which
we address in the following part. Additional external shear and normal
pressures (see \ref{sec:pressure} below) $p_s$ and $p_n$, respectively, can be
included as additional contributions $-p_s$ on the right hand side of the
tangential stress equilibrium (i) and $-p_n$ on the right hand side of the
normal stress equilibrium (iii).  Instead we will include $E_\sigma$ in the
free energy and derive matching conditions from the variational calculus in
Appendix \ref{app:match}.

\subsubsection{Hookean elasticity and alternative 
   constitutive relations}

The shape equations in the form 
(\ref{eqn:shape-equations}) are still  independent 
of the elastic material law. They only contain stress and moment 
equilibria and 
geometrical relations and no information on the 
 constitutive relation characterizing the material. 
The constitutive relations are needed 
to close the shape equations as described in the main text.
We use a nonlinear Hookean elasticity with  
constitutive equations \eqref{eqn:constitutive-laws}, 
which derive from the Hookean elastic energy \eqref{eqn:hooke-energy}. 

For large stretching tensions $\sigma$ such that
  $\sigma > Y_\mathrm{2D}$ ($H <   L_\sigma$) and 
$\sigma > \gamma$, the stretching strains 
$e_{s,\phi} = (\lambda_{s,\phi}^2-1)/2$ are no longer small 
and the linear 
approximation  $e_{s,\phi} \approx \lambda_{s,\phi} - 1$
breaks down. Then the essentially linear 
Hookean constitutive relation \eqref{eqn:constitutive-laws}
is no longer valid and has to be replaced by more specific  
nonlinear hyperelastic  material laws, such as Mooney-Rivlin or 
 Skalak laws \cite{Libai1998,Barthes-Biesel2002}.
For the Mooney-Rivlin  law it has been explicitly demonstrated in Ref.\ 
\citenum{hege2017elastometry} how this constitutive relation 
can be used to close the shape equations (\ref{eqn:shape-equations}).
Because the shape equations \eqref{eqn:shape-equations} are
closed by eliminating $\lambda_s$ and $\tau_\phi$ by using the 
constitutive relations for stresses and strains, nonlinear 
relations such as Mooney-Rivlin laws are considerably more difficult 
to handle and lead to a larger computational cost \cite{hege2017elastometry}.

\begin{figure*}
\centering
\includegraphics[width=\linewidth]{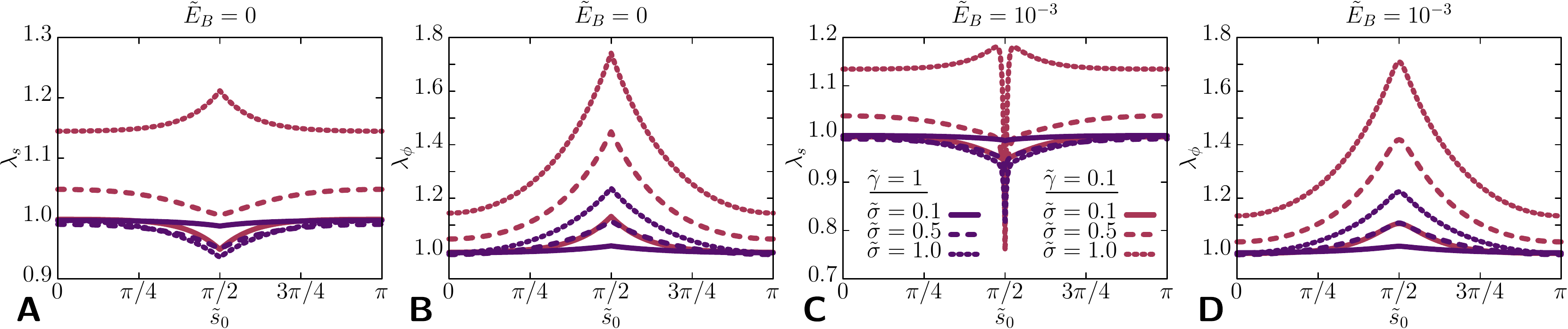}
\caption{
  Stretching ratios $\lambda_{s}$ (A,C) and $\lambda_\phi$ (B,D)
   along the capsule contours 
for $\tilde{E}_B=0$ (A,B) and $\tilde{E}_B=10^{-3}$ (C,D)
in the elastic  regime  $\gamma /Y_\mathrm{2D} = \tilde{\gamma} =1/10$ 
(red) and the  crossover regime $\tilde{\gamma}= 1$ (purple) 
and for different tensile stresses 
  $\tilde{\sigma}= \sigma /Y_\mathrm{2D}$ for the symmetric case. 
Whereas meridional strains $|\lambda_s-1|<0.2$ remain small in (A,C), 
circumferential strains $|\lambda_\phi-1|$ can become 
large at  the liquid-liquid interface, where tension is exerted. 
In the elastic 
 regime $\tilde{\gamma} = 1/10$, values $\lambda_\phi\simeq 1.8$ 
are reached for $\tilde{\sigma}=1$. 
}
\label{fig:lambdas}
\end{figure*}

In Fig.\ \ref{fig:lambdas},
 we investigate in more detail whether nonlinear effects can play 
a role for capsules stretched with tensile 
stresses  $\sigma \le Y_\mathrm{2D}$, which are realistic 
for typical experimental situations and also used for the shape
evolutions in Fig.\ \ref{fig:num-solA}. We find that 
meridional strains remain small, whereas circumferential 
stresses can indeed become large locally in the vicinity 
of the liquid-liquid interface
 (reaching values  $\lambda_\phi \simeq 1.8$
for  $\sigma = Y_\mathrm{2D}$). 
This will lead to a further stretching of the capsule shape 
if, for example, a more realistic 
Mooney-Rivlin elastic law is used which 
displays a softening for large strains \cite{Barthes-Biesel2002}.
We do not expect, however, that any of our results 
will qualitatively change if more realistic nonlinear elastic laws 
are used in this regime.

Another problem arises if compressive stresses $\tau_s<0$ or $\tau_\phi<0$ 
occur, which can give rise to wrinkle formation 
and require  a different effective constitutive relation
in the compressed region
\cite{Knoche2013,hege2017elastometry}. 
We have checked explicity that 
such compressive stresses do not occur 
 for capsules stretched at a liquid-liquid interface, i.e., $\sigma>0$.
For compressive tensions $\sigma<0$, compressive stresses $\tau_\phi<0$ 
can indeed occur, such that wrinkles in $s$-direction could form in the
vicinity  of  the AB-interface. We do not include wrinkle formation 
in our analysis as we focus on tensile stresses $\sigma>0$ 
throughout this paper.

\subsubsection{Pressure and  external forces}
\label{sec:pressure}

Assuming an incompressible liquid within the capsule and an impermeable
membrane leads to conservation of the capsule volume
(\ref{eqn:vol}).
The volume constraint $V= V_0=\mathrm{const}$
is satisfied by adjusting the hydrostatic
pressure $p_0$, which serves as a Lagrange parameter. 
In practice, the volume constraint is 
 realized by including it in the shooting method, i.e., 
by using $p_0$ as shooting parameter to obtain a given volume $V_0$ 
at the end of the integration along the contour.

If $p_0$ is not interpreted as Lagrange parameter 
we have  actual pressure control. 
Then  the pressure $p_0$  is prescribed
and  determines the 
capsule volume $V=V(p_0)$. 
 Solutions for pressurized capsules are
numerically simpler to obtain and possibly exhibit more diverse shapes due to
the lacking volume constraint. 
Another possibility, which is intermediate between pure pressure and 
pure volume control,  is osmotic pressure control \cite{Knoche2014osmotic}. 
Then the osmotic pressure inside the capsule becomes a function of 
the capsule volume $p_0=p_0(V)$, the shape of which depends on the osmolyte 
concentration.  
We restrict ourselves here to capsules
with a constant volume in which the hydrostatic pressure adapts accordingly.

External forces arises from the surface tension of the 
AB-interface which acts along the contact line 
$z=0$ at arc length $s_0=\ell$ around the capsule. 
This can be interpreted as a  point load  on
the capsule contour, i.e., a localized  force density (\ref{eqn:fsigma}).

The force density could be included as a pressure contribution 
into the shape equations \eqref{eqn:shape-equations}
leading to normal and tangential
pressure contributions
\begin{align}
\begin{split}
p_n &= \vec{f}_\sigma \cdot \vec{n} = \sigma \delta (s_0-\ell) \sin\psi, \\
p_s &= -\vec{f}_\sigma \cdot \vec{e}_s = -\sigma \delta (s_0-\ell) \cos\psi.
\end{split}
\label{eqn:ext-forces}
\end{align}
The total normal pressure is then given by $p = p_0 + p_n$.
 We will rather include the point loads 
by deriving   adequate  matching  conditions 
for the forces $\tau_s$ and $q$   
at $z=0$ or $s_0=\ell$  from variation of the energy 
(\ref{eqn:Gvar}) 
in the following section \ref{app:match}.

\subsection{Matching conditions at the liquid-liquid interface}
\label{app:match}

At the liquid-liquid interface at $z=0$ and $s_0=\ell$ 
the shape equations (\ref{eqn:shape-equations}) have to be complemented 
by matching  conditions.  
We argued in the main text
 that six matching conditions are needed for  a 
Hookean shell, whose shape is obtained  by the full set 
 (\ref{eqn:shape-equations}) of six shape equations, whereas only 
four matching conditions are needed for a Hookean membrane, which is 
described by four shape equations
 ($m_s=0$ and $q=0$ in eqs.\ (\ref{eqn:shape-equations})).
 Moreover, we need to determine the parameter $\ell$ 
itself by energy minimization  in the
general asymmetric case. For the symmetric case
$\gamma_\mathrm{A}=\gamma_\mathrm{B}$ we have 
$\ell = L_0/2$ by symmetry. 
The capsule consists of two solution branches  $z>0$ and $z<0$,
for which we will use superscripts $+$ and $-$, respectively, in the following
to formulate the matching conditions.

 We will derive  all matching  conditions 
 from continuity conditions for intact closed shells and 
variation of the energy 
(\ref{eqn:Gvar}) with 
respect to the position of the three phase contact line
at $z=0$ and $s_0=\ell$.

We have three 
 obvious matching conditions  for $z$ and $r$,
\begin{equation}
z^+(\ell)=z^-(\ell)=0~~\mbox{and}~~  r^+(\ell)=r^-(\ell)\equiv r(\ell),
\label{eqn:r_cont}
\end{equation}
from requiring a closed capsule. 
The two matching conditions for $z$ 
 fix the shooting parameters $z(0)$ and
$z(L_0)$.
From continuity of $r$ also the continuity of $\lambda_\phi = r/r_0$ 
follows immediately. 
For a shell with bending rigidity we also have the 
fourth matching  condition
\begin{equation}
\psi^+(\ell) = \psi^-(\ell)
\label{eqn:psi_cont}
\end{equation}
 because a 
kink in the capsule shell costs an infinite bending energy. 
Membranes and droplets, however, will exhibit such kinks 
and $\psi^+(\ell)$ and $\psi^-(\ell)$ can freely adjust. 
From continuity of $r$ and $\psi$, also the continuity of
the curvature  $\kappa_\phi = \sin\psi / r$  follows.

All remaining matching conditions are derived based on  the variational 
calculus introduced in Ref.\ \citenum{Knoche2011} by
minimizing the total free energy  
$G= \int w(s_0) \mathrm{d}A_0 - p_0V-\sigma \pi r^2(\ell)$
or (using eq.\ (\ref{eqn:vol})) 
\begin{align}
G &=   \int_0^{\ell} \mathrm{d}s_0  g^- + \int_{\ell}^{L_0} \mathrm{d}s_0 g^+
    - \sigma \pi r(\ell)^2,
~~\mbox{with}
\nonumber\\
g^\pm &\equiv  2\pi r_0 w^\pm
    - p_0 \pi (r^\pm)^2  \lambda_s^\pm\sin\psi^\pm,
\label{eqn:min-energy}
\end{align}
where $w^\pm(s_0)$ are the Hookean energy densities
  (\ref{eqn:hooke-energy})
of the upper and  lower parts of the capsule, which have identical 
elastic constants $Y_\mathrm{2D}$, $\nu_\mathrm{2D}$, and $E_\mathrm{B}$
but differ in  
their surface  tension contributions with $\gamma_A$ in the lower ($-$) 
and $\gamma_B$ in the upper ($+$) part;
 the energy $E_\sigma=-\sigma \pi r^2(\ell)$ is the potential  for the point 
force $\vec{f}_\sigma$ from eq.\ (\ref{eqn:fsigma}).
The free energy  $G$ has to be extremized with respect to the functions 
$r^\pm(s_0)$  and $\psi^\pm(s_0)$  as well as with respect to the 
location $\ell$ of the liquid-liquid interface. 
The total variations  $\delta r \equiv (r +\delta r)(\ell+\delta \ell)
 -r(\ell) = \delta r(\ell) + r'(\ell) \delta \ell$ 
and $\delta \psi \equiv (\psi +\delta \psi)(\ell+\delta \ell)
 -\psi(\ell) = \delta \psi(\ell) + \psi'(\ell) \delta \ell$ 
at the variable interface position $\ell$
have to fulfill the continuity conditions 
$\delta r^- = \delta r^+$ and, for a shell, 
$\delta \psi^-= \delta \psi^+$.
 
 Extremizing with respect to the functions 
$r^\pm(s_0)$  and $\psi^\pm(s_0)$  leads to the same Euler-Lagrange equations 
(\ref{eqn:EL}) or (\ref{eqn:equilibrium_var}), which hold 
 both for the upper (+) and 
lower (-) parts.
In variational calculus with variable functions at the boundary
at $s_0=\ell$,
each continuity condition at the boundary 
entails a corresponding Weierstrass-Erdmann
condition.
 Equating all  boundary terms $\propto \delta r^+=\delta r^-$ in the variation 
$\delta G$ to zero we obtain the  Weierstrass-Erdmann condition for $r$
\begin{equation}
  0=  g^-_{r'}\big|_\ell  -  g^+_{r'}\big|_\ell 
      + \frac{\mathrm{d}}{\mathrm{d} r(\ell)}E_\sigma
    =  g^-_{r'}\big|_\ell -  g^+_{r'}\big|_\ell  -r(\ell) \sigma.
\label{eqn:WEr}
\end{equation}
Likewise, equating all
  boundary terms $\propto \delta \psi^+=\delta \psi^-$ in the variation 
$\delta G$ to zero for shells, we find the  Weierstrass-Erdmann condition 
for $\psi$
\begin{equation} 
  0=  g^-_{\psi'}\big|_\ell -  g^+_{\psi'}\big|_\ell  
    + \frac{\mathrm{d}}{\mathrm{d} \psi(\ell)}E_\sigma
    =  g^-_{\psi'}\big|_\ell -  g^+_{\psi'}\big|_\ell.
\label{eqn:WEpsi}
\end{equation}
Moreover, 
 because of  the variable interface position 
we also have to equate
 all  boundary terms $\propto \delta \ell$ in the variation 
$\delta G$ to zero (at constant volume), which 
gives an 
additional transversality condition
\begin{equation}
  \left(g^+- (r^+)' g^+_{r'} - (\psi^+)' g^+_{\psi'}\right)\big|_\ell = 
    \left(g^-- (r^-)' g^-_{r'} - (\psi^-)' g^-_{\psi'}\right)\big|_\ell,
\label{eqn:Trans}
\end{equation}
where the $\psi$-terms are only present for shells.

As derived above in eq.\ (\ref{eqn:q0}), 
variation of $G$  gives the additional 
algebraic relation
\begin{equation}
    q\cos\psi +\tau_s \sin\psi = \frac{1}{2} p_0 r,
\label{eqn:q}
\end{equation}
for $s_0 \neq \ell$, 
which can replace the shape equation for
 $q$ in   \eqref{eqn:shape-equations}. 
This algebraic equation is equivalent
to force equilibrium  in $z$-direction 
for any  part of the capsule from the 
lower apex up to arc length $s$.
In equilibrium, the 
 total force $F_z$ 
in axial $z$-direction that 
is acting on the lower part of the capsule from the 
lower apex up to arc length $s$ vanishes 
because of axial symmetry and the absence of external 
forces such as gravity acting in $z$-direction
 \cite{Boltz2015,Boltz2016},
\begin{align}
 0= F_z(s) &=  2\pi r (q \cos\psi + \tau_s \sin\psi) 
\nonumber\\
   & ~~-
    2\pi   \int_0^s \mathrm{d}\tilde{s} r((p_0+p_n)\cos\psi + p_s \sin\psi)
\nonumber\\
   &= 2\pi r (q \cos\psi + \tau_s \sin\psi) - \pi p_0 r^2,
\label{eqn:U}
\end{align}
where we used eq.\ (\ref{eqn:ext-forces}) for $p_n$ and $p_s$ to obtain 
the last equality. 
 The first term is the 
total interfacial  force in $z$-direction integrated 
along the rim of length $2\pi r$, the second part 
the total pressure force acting on the area of the lower capsule part
in $z$-direction. There are no 
additional forces in $z$-direction from the liquid-liquid interface
such that both contributions must cancel for all $s$, also 
at the liquid-liquid interface at $s=\ell$.

Equation (\ref{eqn:q}) also implies that $q=0$ at the apices where
$r=0$ and $\psi = 0$ or $\pi$, which justifies the 
boundary conditions $q(0)=q(L_0)=0$. 
At the interface at $s_0=\ell$ the right hand side of eq.\ (\ref{eqn:q}) 
is continuous because $r$ is continuous, 
$r^+(\ell)=r^-(\ell)= r(\ell)$
such that 
\begin{align}
  & \left(q^+\cos\psi^+ +\tau_s^+ \sin\psi^+\right)\big|_\ell 
 =  \left( q^-\cos\psi^- +\tau_s^- \sin\psi^-\right)\big|_\ell
= \frac{1}{2} p_0 r(\ell).
\label{eqn:q_match}
\end{align}
This is equivalent to $F_z^+(\ell) = F_z^-(\ell) =0$, 
which holds  because the liquid-liquid interface 
exerts no force in $z$-direction. 
The first equality in eq.\ (\ref{eqn:q_match}) describes 
a force equilibrium in the $z$-direction
 at the contact line between capsule and liquid 
interfacial tensions and, thus, follows from 
a Neumann triangle construction (see also 
Fig.\ \ref{fig:num-solB}).
The liquid-liquid interface is not exerting interfacial 
forces in the $z$-direction and, 
thus, $\sigma$ does not enter eq.\  (\ref{eqn:q_match}); discontinuities 
in the normal shear force density $q$  and in the tangential 
force density $\tau_s$ have to cancel in $z$-direction. 

The second equality in eq.\ (\ref{eqn:q_match}) shows 
that the continuity of $r^+(\ell)=r^-(\ell)= r(\ell)$ is actually 
{\it equivalent} to the Neumann tension force equilibrium in $z$-direction
at the contact line because continuity of $r$ leads to continuity 
of the pressure force $\pi p_0 r^2$ in $z$-direction  and 
interfacial and pressure forces are always opposite and equal
according to the 
 relation  $F_z(s)=0$ for the total force. 
Therefore, using both matching conditions (\ref{eqn:q_match}) and 
 $r^+(\ell)=r^-(\ell)= r(\ell)$ always leads to an over-determined 
set of matching conditions. 

Using the definitions (\ref{eqn:gdef}) in  the 
Weierstrass-Erdmann condition (\ref{eqn:WEr}) from the 
variation $\delta r^- = \delta r^+$ at the boundary gives
\begin{align}
  &\left(q^+\sin\psi^+ -\tau_s^+ \cos\psi^+\right)\big|_\ell   
  = \left(q^-\sin\psi^- -\tau_s^- \cos\psi^-\right)\big|_\ell +\sigma,
\label{eqn:Young}
\end{align}
which describes  the  force equilibrium in the radial direction
 at the contact line between capsule and liquid 
interfacial tensions  and, thus, also follows from 
the Neumann triangle construction (see also 
Fig.\ \ref{fig:num-solB}).
 Discontinuities 
in the normal shear force density $q$  and in the tangential 
force density $\tau_s$ in $r$-direction 
have to cancel with the tension $\sigma$, which also acts  
in $r$-direction. 

Likewise, 
using the definitions (\ref{eqn:gdef}) in the 
Weierstrass-Erdmann condition (\ref{eqn:WEpsi}) from the 
variation $\delta \psi^- = \delta \psi^+$ at the boundary gives
\begin{equation}
  m_s^+(\ell)=m_s^-(\ell),
\label{eqn:m_cont}
\end{equation}
which holds for shells and describes the moment equilibrium at the   
contact line.
Continuity of $m_s$ and $\kappa_\phi$ (see above) also entails continuity 
of $\kappa_s$ and $m_\phi$ and, thus, of the entire bending 
energy density.

Inserting the definitions (\ref{eqn:gdef}) in the 
transversality  condition (\ref{eqn:Trans}) and also utilizing
the continuity of $r$ (and $m_s$ and $\kappa_s$ for shells)  and 
the first equation from the force  equilibrium in the $z$-direction
eq.\ (\ref{eqn:q_match}) finally we find
\begin{equation}
  \left(w^+ - \lambda_s^+\lambda_\phi^+\tau_s^+\right)\Big|_\ell =  
   \left( w^- 
  - \lambda_s^-\lambda_\phi^-\tau_s^-\right)\Big|_\ell.
\label{eqn:transversality}
\end{equation}
Because the bending energy part of $w$ is continuous, it follows 
that the discontinuity in $w$ across 
the interface is exactly due to the discontinuity in $\gamma$, 
which jumps from $\gamma^-=\gamma_\mathrm{A}$  for $z<0$ to 
$\gamma^+=\gamma_\mathrm{B}$  for $z>0$.
This means, in turn, that the stretching elasticity part 
of $w$ is also continuous. Because $\lambda_\phi$ is continuous (see above), 
also $\lambda_s$ is continuous and, thus, the transversality condition 
\eqref{eqn:transversality} means that 
the elastic parts of the tensions $\tau_s - \gamma$ and 
$\tau_\phi - \gamma$ have to be  continuous across the interface,
\begin{equation}
 \left( \tau_s^+-\gamma^+\right)\big|_\ell = 
\left( \tau_s^--\gamma^-\right)\big|_\ell.
\label{eqn:transversality2}
\end{equation}
In our numerical treatment, we will not employ the transversality condition in
the forms (\ref{eqn:transversality}) or (\ref{eqn:transversality2})
but prefer to numerically minimize the
total energy with respect to $\ell$
(changing the interface arc length $\ell$ for this minimization
requires a re-meshing in the multiple shooting method that 
is explained in the next section).

\subsection{Numerical method/precision}
\label{app:shooting}

\begin{figure}
\centering
\includegraphics[width=\linewidth]{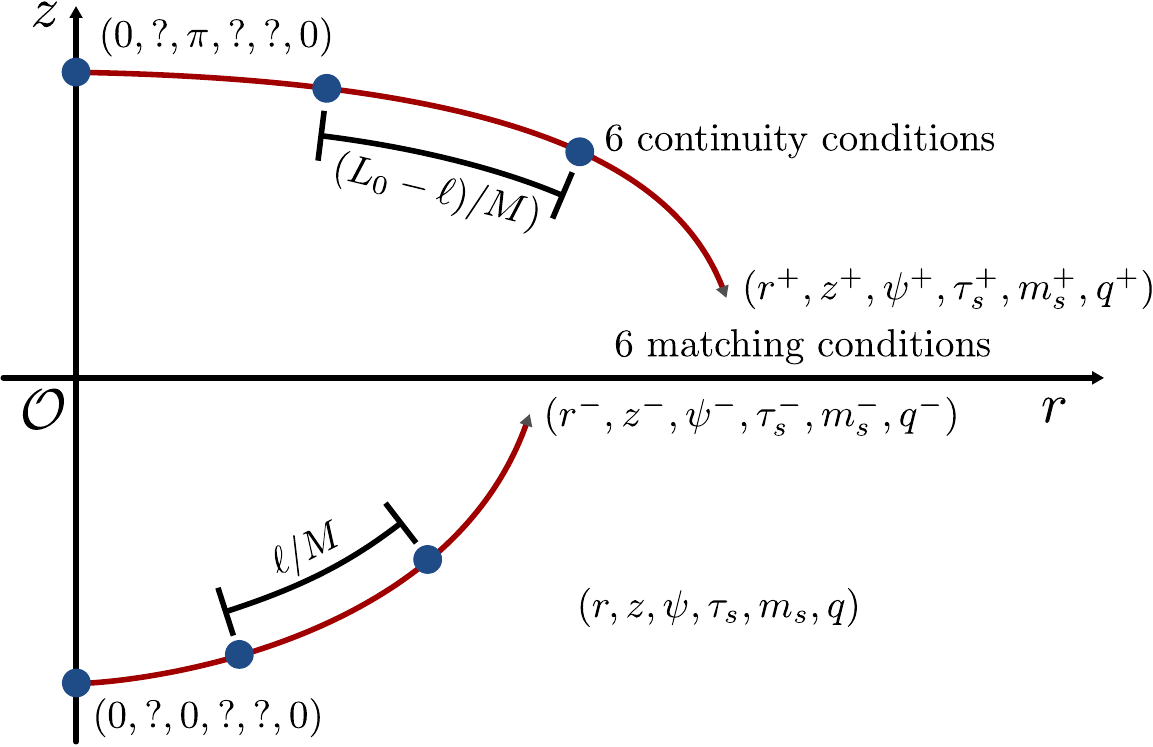}
\caption
  {Illustration of the multiple shooting method
  applied to numerically solve the shape equations.  The $r$-$z$-plane is
  separated by a liquid-liquid interface at $z=0$ arising from two different
  liquid phases $\mathrm{A}$ and $\mathrm{B}$ residing in the lower
  and  upper half-space, respectively.  
  The lower and upper branches are integrated
  from the apices on the $z$-axis to the interface on the $r$-axis, where we
  employ matching conditions corresponding to the different limit cases
  (Hookean membrane and Hookean shell) as discussed in this section.}
\label{fig:scheme-shooting-method}
\end{figure}

To achieve numerical stability we 
employ a multiple shooting method (Fig.\ \ref{fig:scheme-shooting-method}),
 where we 
subdivide each solution branch in $M$
segments.  At each intermediate point we gain six shooting parameters and six
continuity conditions (for the general shell case).  
The $(-)$-branch is integrated counter-clockwise over
the interval $[0,\ell]$ starting at $s_0=0$ and ending at $s_0=\ell$, the
$(+)$-branch is integrated clockwise over $[\ell, L_0]$ starting at $s_0=L_0$
and ending at $s_0=\ell$. Both branches thus start at the axis of rotation and
match at the interface.  When the interface arc length $\ell$ is changed, the
segmentation of the shape has to be adapted, such that the $(-)$-branch is
integrated over the intervals $[k\ell/M,(k+1)\ell/M]$ where $k=0,\dots,M-1$,
and the $(+)$-branch is integrated over the intervals
$[L_0-(k+1)(L_0-\ell)/M,L_0-k(L_0-\ell)/M]$. During the minimization
\eqref{eqn:min-energy} we iteratively change the segmentation of the shape,
i.e., change $\ell$ until the total free energy $G$
reaches its minimum.

For the multiple shooting we employ a least square minimization technique that
minimizes the distances between the individual shape segments, the matching
conditions at the interface, as well as the deviation from the target volume
$V_0$.  We therefore compute the Jacobian in order to obtain descent
directions, that we follow iteratively. We assume that this algorithm is
converged, if the Euclidean norm of the (dimensionless) 
residual vector falls below $\epsilon
= 10^{-8}$. This way we obtain valid solutions of the shape equations for a
given branch junction length $\ell$. Now, to obtain asymmetric shapes we need
to minimize \eqref{eqn:min-energy} by changing $\ell$ iteratively. 
Therefore, we use a one-dimensional method (similar method as described above), 
which terminates on falling below a step length 
$\Delta \ell_\mathrm{step} = 10^{-8}R_0$.

\section{Shell, membrane and droplet regime}

In the following, we discuss numerical solutions 
of  the shape equations for the general asymmetric case 
for  three different types of interface elasticity:
liquid lenses,  lens-shaped capsules with  
a Hookean membrane (elastic lenses),
and capsules with  Hookean shell elasticity as 
shown in  Fig.\ \ref{fig:num-solA}.
We specify shape equations and the matching conditions
for the three different regimes.

\subsection{Liquid lens}

The liquid lens is the simplest case discussed in this section, as it is
obtained in the absence of any elastic tension, 
  $Y_\mathrm{2D} \approx 0$ ($\tau_s=\tau_\phi=\gamma$)
and $E_\mathrm{B}\approx 0$ 
($q=0$ and $m_s=m_\phi=0$). 
For this liquid surface there is no reference shape  and
  surface tension and incompressibility determine the droplet shape.
From the local force balance condition $p_0\mathrm{d}V = \sigma\mathrm{d}A$,
we obtain the Laplace-Young equation,
\begin{align}
p_0 = \gamma (\kappa_{s} + \kappa_{\phi}),
\label{eqn:laplace-young-non}
\end{align}
which can be recast as a set of three shape equations
\begin{align}
\begin{split}
r^{'}(s_0) &= \cos\psi, \hspace*{1em} z^{'}(s_0) = \sin\psi, \\
\psi^{'}(s_0) &= p_0/\gamma - \sin\psi/r,
\end{split} 
\label{eqn:laplace-young-se}
\end{align}
using cylindrical parametrization.  This set of equations also directly
derives from \eqref{eqn:shape-equations} by employing the limits given
above. The resulting shapes have \emph{constant} mean curvature according 
to the Laplace equation \eqref{eqn:laplace-young-non}
(or according to $\kappa_s + \kappa_\phi = \psi^{'}+ \sin\psi / r = {\rm const}$ 
in \eqref{eqn:laplace-young-se}), which 
only allows lens shapes  that are composed of two spherical caps with 
the same radius $R$.

For the numerical determination of the 
shape of the liquid lens we shoot from both apices with 
boundary conditions $r(0)=r(L_0)=0$ and $\psi(0)=\pi-\psi(L_0)=0$.
We will determine two shooting parameters  $z(0)$ and
$z(L_0)$ such that $z^+(\ell)=z^-(\ell)=0$. 
Because we have three shape equations \eqref{eqn:laplace-young-non},
there are no free shooting parameters left. 

There are, however,  three  matching conditions at the 
AB-interface at $s_0=\ell$: the 
Neumann triangle  conditions (\ref{eqn:q_match}) and
(\ref{eqn:Young}) for force equilibrium in $z$- and $r$-direction,
respectively, which become ($q=0$, $\tau_s^-=\tau_\phi^-=\gamma_\mathrm{A}$,
$\tau_s^+=\tau_\phi^+=\gamma_\mathrm{B}$)
\begin{align}
\begin{split}
   f_r &\equiv \gamma_\mathrm{A} \cos\psi^-(\ell) -\gamma_\mathrm{B}
   \cos\psi^+(\ell)  -  \sigma =0\\
  f_z &\equiv \gamma_\mathrm{A} \sin\psi^-(\ell) -\gamma_\mathrm{B}
   \sin\psi^+(\ell) =0
\end{split}
\label{eqn:young-equation-liquid}
\end{align}
and the continuity condition $r^-(\ell)= r^+(\ell)$.
As shown 
in section \ref{app:match}, this continuity equation is actually 
equivalent to the Neumann condition $f_z=0$ for force equilibrium in
$z$-direction.
For the liquid lens,
 there is no reference shape, and the matching conditions 
have to be used to determine the arc lengths 
$L_0-\ell$ and  $\ell$ of the upper and lower part.
We thus have two independent matching conditions for two unknown parameters 
$\ell$ and $L_0$. 
Numerically, we use the full over-determined set of three 
matching conditions because we find that 
this leads to faster convergence of the shooting method. 
We note that liquid lenses exhibit a kink at the
interface, i.e., there is no continuity of $\psi$ at the AB-interface.
We also note that, because there is no elastic energy or reference shape,
 $L_0$ is not fixed 
beforehand by the reference shape and $\ell$ does not have to be 
determined from energy minimization as will be the case for Hookean 
membranes and shells.

\subsection{Elastic lens}

In case of finite stretching resistance $Y_\mathrm{2D}$, 
but vanishing bending rigidity $E_\mathrm{B}\approx 0$, i.e., 
vanishing shell thickness $H\approx 0$,
we have an elastic lens with Hookean membrane elasticity
with vanishing bending moments $m_s=m_\phi=0$ and vanishing transverse
shear stress $q=0$.
The system \eqref{eqn:shape-equations} reduces to 
four coupled nonlinear differential equations
\begin{align}
\begin{split}
r^{'}(s_0) &= \lambda_s\cos\psi,\hspace*{1em} 
   z^{'}(s_0) = \lambda_s\sin\psi,  \\
\psi^{'}(s_0) &= \lambda_s(p  - \kappa_\phi\tau_\phi)/\tau_s, \\
\tau_s^{'}(s_0) &=  \lambda_s\left(\frac{\tau_\phi -\tau_s}{r}
 \,\cos\psi+p_s\right).
\end{split}
\label{eqn:thin-shape-equations}
\end{align}
Shapes are still
similar  to liquid lenses, and there is a 
 kink at the AB-interface because 
 there is no continuity of $\psi$.

For the numerical determination of the 
elastic lens shape we shoot from both apices with 
boundary conditions $r(0)=r(L_0)=0$ and $\psi(0)=\pi-\psi(L_0)=0$.
The two shooting parameters  $z(0)$ and
$z(L_0)$ are determined from  $z^+(\ell)=z^-(\ell)=0$. 
Because we have four  shape equations \eqref{eqn:thin-shape-equations},
there are two free shooting parameters $\tau_s(0)$ and $\tau_s(L_0)$ left,
which have to be determined from matching conditions at 
the AB-interface at $s_0=\ell$. 

As for a liquid lens,  there are three  matching conditions: the 
Neumann triangle conditions (\ref{eqn:q_match}) and
(\ref{eqn:Young})  for force equilibrium in $z$- and $r$-direction,
respectively, which become ($q=0$)
\begin{align}
\begin{split}
f_r &\equiv \tau_s^-(\ell)\cos\psi^-(\ell) - \tau_s^+(\ell)\cos\psi^+(\ell) 
 -\sigma = 0 \\
f_z &\equiv \tau_s^-(\ell)\sin\psi^-(\ell) - \tau_s^+(\ell)\sin\psi^+(\ell) = 0,
\end{split}
\label{eqn:young-equation}
\end{align}
and the continuity condition $r^-(\ell)= r^+(\ell)$.
Again, this continuity equation is actually 
equivalent to the Neumann condition $f_z=0$ for force equilibrium in
$z$-direction as shown in section \ref{app:match}.
Two independent   matching conditions can be 
 used to determine the two shooting
parameters $\tau_s(0)$ and $\tau_s(L_0)$. Again, we use the full
over-determined set of three matching conditions, which 
leads to faster convergence of the shooting method.
Finally, the arc length position $\ell$ of the AB-interface 
is determined by total energy minimization.

In the symmetric case, 
where $\gamma_\mathrm{A} = \gamma_\mathrm{B}$
and $\psi^-(\ell)=\pi-\psi^+(\ell)$, we find $\tau_s^-(\ell)=\tau_s^+(\ell)$. 
Note that the tensions $\tau_s^-$ and $\tau_s^+$ include the
liquid interface tensions $\gamma_\mathrm{A}$ and $\gamma_\mathrm{B}$ 
in addition to elastic contributions. 
In the symmetric case, the shooting
parameters $\tau_s$ 
at the upper and lower apices are identical and thereby
only one parameter is left.  Symmetry with respect to the liquid-liquid
interface also inherently satisfies the conditions $r^+(\ell) - r^-(\ell) = 0$
and $f_z = 0$ such that we are left with exactly one matching 
conditions $f_r=0$ from the Neumann condition
(\ref{eqn:Young})  for force equilibrium in  $r$-direction.

Numerically calculated shapes of elastic lenses are shown in Fig.\ 
\ref{fig:num-solA} both for tensile  ($\sigma >0$) and 
contractile ($\sigma <0$) interfacial tensions.

\subsection{Elastic shells}

Finally, we incorporate the effect of a finite bending rigidity. 
Then we have the full set \eqref{eqn:shape-equations}  of six shape 
equations. 
We shoot  from both apices with 
boundary conditions $r(0)=r(L_0)=0$,  $\psi(0)=\pi-\psi(L_0)=0$ and 
 $q(0)=q(L_0)=0$.
The two shooting parameters  $z(0)$ and
$z(L_0)$ are determined from  $z^+(\ell)=z^-(\ell)=0$.
Now we have four free shooting parameters $\tau_s(0)$, $\tau_s(L_0)$,
$m_s(0)$, and $m_s(L_0)$ left,
which have to be determined from matching conditions at 
the AB-interface at $s_0=\ell$.

There are five matching conditions at the AB-interface:
the continuity conditions $r^-(\ell)= r^+(\ell)$ and also 
$\psi^+(\ell) = \psi^-(\ell)$  (see eq.\ (\ref{eqn:psi_cont})) 
because kinks in the capsule shell are now suppressed by 
bending energy. Moreover, we also have the continuity condition 
$m_s^+(\ell)=m_s^-(\ell)$ (see eq.\ (\ref{eqn:m_cont})) from the 
moment equilibrium at the AB-interface and 
two Neumann conditions (\ref{eqn:q_match}) and
(\ref{eqn:Young})  for force equilibrium in $z$- and $r$-direction.
The two  Neumann conditions can be rearranged into 
direct jump conditions for $\tau_s$ and $q$ at the AB-interface,
\begin{align}
\begin{split}
 q^+(\ell) - q^-(\ell) &= \sigma\sin\psi^-(\ell)
\\
\tau_s^+(\ell) - \tau_s^-(\ell) &= -\sigma\cos\psi^-(\ell).
\end{split}
\label{eqn:jump-conditions}
\end{align}
Again, one Neumann condition  is actually 
equivalent to the continuity condition $r^-(\ell)= r^+(\ell)$
as shown in section \ref{app:match},
such that we have a 
set of  four independent  matching conditions 
for four shooting parameters.
Numerically,  we achieve faster convergence
employing the over-determined set of all  five matching conditions. 
The arc length position $\ell$ of the AB-interface 
is determined by total energy minimization. 

Numerically calculated shapes for elastic shells are also 
shown in Fig.\ 
\ref{fig:num-solA} both for tensile  ($\sigma >0$) and 
contractile ($\sigma <0$) interfacial tensions.

\section{Analytical description of adsorbed shapes}

As it is evident from eq.\ \eqref{eqn:laplace-young-se} liquid lens
shapes exhibit constant curvature and, thus, can be constructed from
spherical caps. In this section, we use this fact to obtain an exact
analytical result for the contact angle and the height of liquid lenses. 
We generalize this approach to elastic lenses by taking also elastic
stresses into account and obtain an approximative theory, which gives
the contact angle and the height of an elastic lens by solving 
numerically a single algebraic equation. Finally, these theoretical
results enable a determination of Young's modulus of an elastic lens
by a \textit{single} measurement of the height \textit{or} the contact angle. 
Furthermore, we give an
 approximative description of the rounding at the contact 
at finite bending rigidity, where the shapes are more discus-
 than lens-shaped, 
that is conceptually analogous to the Pogorelov approach to buckling.

\subsection{Liquid lens}
\label{sec:liquidlens_app}

\begin{figure*}[t]
\centering
\includegraphics[width=\linewidth]{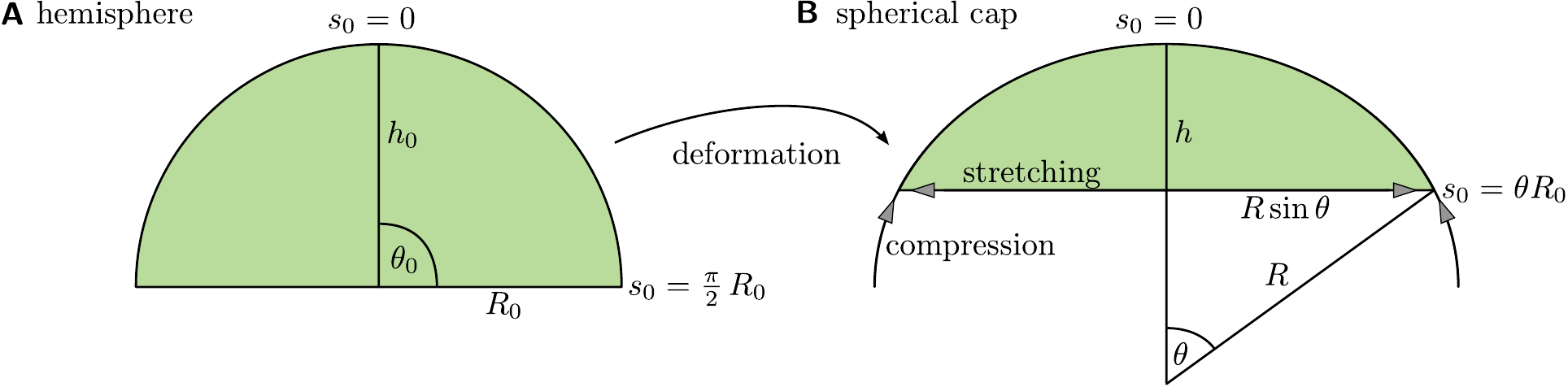}
\caption{
  Deforming a hemispherical rest shape  (A) into a 
  spherical cap (B). 
  The resulting spherical cap has constant tension $\tau_s =\mathrm{const}$ in
  meridional direction, constant tension $\tau_\phi=\mathrm{const}$ in
  circumferential direction, but tensions are anisotropic, i.e., $\tau_s \neq
  \tau_\phi$  violating the force balance conditions
  \protect\eqref{eqn:equilibrium_var} (i). 
  We conclude that spherical caps with hemispherical rest shape 
  exhibit inhomogeneous stresses in
  order to fulfill \protect\eqref{eqn:equilibrium_var} (i). }
\label{fig:spherical_caps}
\end{figure*}

Lenticular liquid 
shapes have been discussed in literature both at a liquid-liquid 
interface \cite{Princen1963} and on a solid substrate
\cite{Widom1995,blecua2006line}, and analytical 
solutions for liquid lenses at liquid-liquid
interfaces can be obtained analogously.  
 We restrict ourselves here to the symmetric case
$\gamma \equiv \gamma_\mathrm{A}=\gamma_\mathrm{B}$. The total
energy of the system is given by
\begin{align}
E &= \gamma A - \sigma A_B,
\label{eqn:free-energy-lens}
\end{align}
where $A$ is the total area of the liquid lens and 
$A_B$  the occupied cross-section area within the liquid-liquid
interface plane.
We use the fact the liquid lens is  composed of two spherical caps
of equal radius $R=2\gamma/p_0$ (for the symmetric case)
according to the Laplace equation \eqref{eqn:laplace-young-non},
and of equal base radius $R_B$.
Then  $A=2\pi(R^2+h^2)$ is the surface of the lens (two spherical caps) and
$A_B=\pi R_B^2$ the occupied cross-section area within the liquid-liquid
interface plane. 
 Here, $h$ denotes the height of the spherical caps,
which is related to $R_B$ and $R$ via $R_B^2 = 2Rh- h^2$, see also
Fig.\ \ref{fig:spherical_caps}(B) for the involved quantities.
In the following, we measure lengths in units of the radius $R_0$
of a spherical droplet of the same volume $V\equiv 4\pi R_0^3/3$
 and thereby introduce
reduced quantities $\tilde{h} \equiv h/R_0$,  $\tilde{R} \equiv R/R_0$,
and $\tilde{V} \equiv V/R_0^3$. 
For a fixed volume, 
\begin{align}
\tilde{V} &= \frac{\pi\tilde{h}}{3}\left(3\tilde{R} - \tilde{h}\right)
    =  \frac{4\pi}{3} = \tilde{V}_0,
\end{align}
the radius $R$ and the height $h$ are related by 
\begin{align}
\tilde{R}(\tilde{h}) &= 
  \frac{1}{3}\left(\frac{2}{\tilde{h}^2}+\tilde{h}\right).
\label{eqn:cap-volume-constraint_app}
\end{align} 
We can now write the energy \eqref{eqn:free-energy-lens} as a function
of the reduced height $\tilde{h}$, minimize 
with respect to the reduced height $\tilde{h}$, and find
the equilibrium height
\begin{align}
\tilde{h} &= \left(\frac{1-\sigma/2\gamma}{1+\sigma/4\gamma}\right)^{1/3}
\label{eqn:h_app}
\end{align}
as a function of $\sigma/\gamma$. 
We  obtain the opening angle $\theta$
from the  geometric relation
\begin{align}
\cos \theta(\tilde{h}) = 1-\frac{\tilde{h}}{\tilde{R}(\tilde{h})}
\label{eqn:wetting-angle_app}
\end{align}
and, finally, the pressure 
 from the Laplace-Young equation
\begin{align}
p_0(\tilde{h}) &= \frac{2\gamma}{\tilde{R}(\tilde{h})} 
  = \frac{6\tilde{h}^2\gamma}{2+\tilde{h}^3}.
\label{eqn:p0_app}
\end{align}
If we use $\tilde{h}$ as a function of $\sigma/\gamma$ from 
eq.\ (\ref{eqn:h_app}) in these relations we find $\theta$ and 
$p_0$ as a function of $\sigma/\gamma$.

The result  (\ref{eqn:h_app})  for $\tilde{h}$ as a function 
of $\sigma/\gamma$ 
can likewise be obtained by starting from the Neumann condition 
 $f_r=0$ in \eqref{eqn:young-equation-liquid},
employing the geometric relation (\ref{eqn:wetting-angle_app}) and 
the volume constraint \eqref{eqn:cap-volume-constraint_app},
\begin{align}
  \sigma = 2\gamma \cos \theta(\tilde{h}) 
     = 2\gamma \left(1-\frac{\tilde{h}}{\tilde{R}(\tilde{h})}\right)
   = 4\gamma  \frac{1-\tilde{h}^3}{2+\tilde{h}^3} 
\label{eqn:Neumann_app}
\end{align}
(for $\theta = \psi^-(\ell)  = \pi - \psi^+(\ell)$ 
and $\gamma =\gamma_\mathrm{A} = \gamma_\mathrm{B}$). 
Solving for  $\tilde{h}$ gives  (\ref{eqn:h_app}).

\subsection{Elastic lens}
\label{sec:elasticlens_app}

For the  symmetric case, we can derive an approximative
analytical solution for the elastic membrane shape based on 
a spherical cap approximation.
We investigate partially spherical solutions, because, intuitively,
the difference between a liquid  droplet and a liquid droplet 
coated with an elastic membrane  should 
be negligible for small interface loads as also 
suggested by the numerically calculated shapes of elastic lenses. 
Force balance for an unloaded spherical cap implies a constant tension 
$\tau_s=\mathrm{const}$ \cite{LL7} such that, additionally,
 $\tau_s = \tau_\phi$ must hold to fulfill
the force balance conditions \eqref{eqn:equilibrium_var} (i).

\begin{figure*}[t]
\centering
\includegraphics[width=\linewidth]{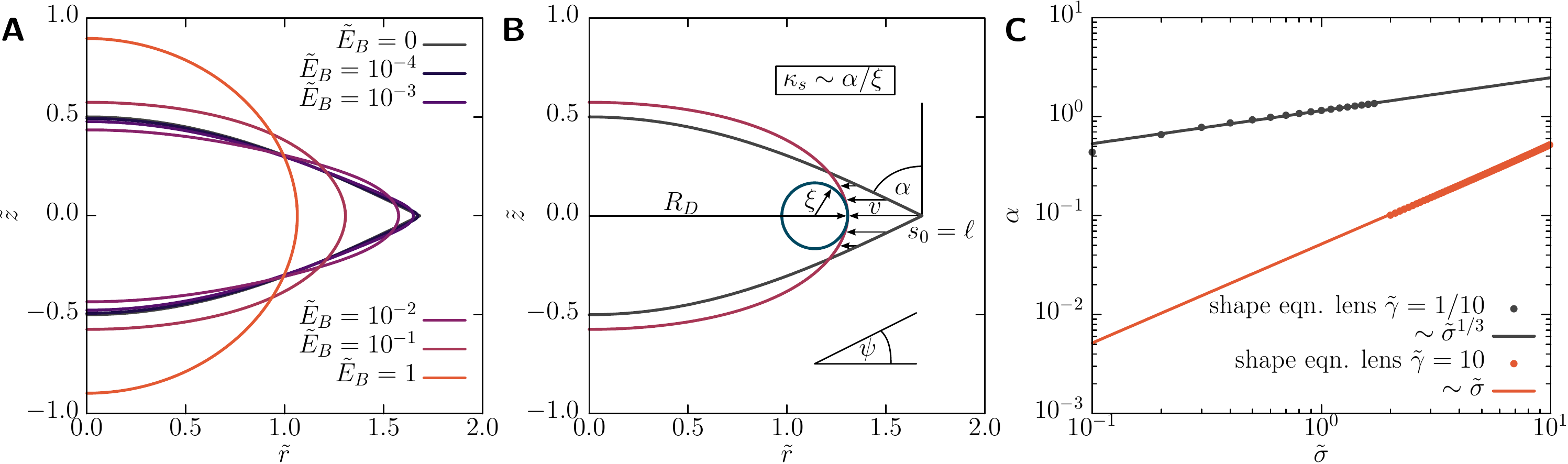}
\caption
  {(A) Shapes with increasing bending modulus.
    The bending modulus
   of the shell's material determines the curvature at the
   liquid-liquid interface at $z=0$ where the interface load $\sigma$ pulls
   the equator outwards. Increasing the bending modulus increases the radius
   of curvature at the interface. Shapes were calculated in the elastic regime
   with $\gamma / Y_\mathrm{2D} = 1/10$.
   (B) Illustration of the geometrical quantities employed
   within the Pogorelov approximation. We assume that the major contribution 
   to the bending energy is located within a small region of size $\xi$ 
   at $s_0=\ell$, i.e., the liquid-liquid interface. When the bending energy
   is switched on, the kink of the elastic lens vanishes and the contact angle 
   $\alpha$ decreases from a finite value to zero. 
  This deformation is quantified
   by the radial displacement field $v(s_0)$. The curvature at the interface is
   accessible via $\kappa_s \sim \alpha/\xi$ employing the length
   scale $\xi$ and
   the contact angle $\alpha$.  
 (C)  Numerical verification of the scaling of the contact angle $\alpha$, 
    see eq.\ (\ref{eqn:alpha-scaling}).
}
\label{fig:spherical-caps-pogorelov}
\end{figure*}

Deforming a hemisphere into a spherical cap by  uniform
strains leads, however, 
 to anisotropic tensions $\tau_s \neq \tau_\phi$. 
The spherical cap has only one free
parameter, which is its reduced height $\tilde{h}$, as it already has been
introduced above in the context of liquid lenses. 
From geometrical arguments (see Fig.\ \ref{fig:spherical_caps})
we can calculate the uniform stretches $\lambda_s$ and 
$\lambda_\phi$  for the 
deformation into a spherical cap,
\begin{align}
\lambda_s &= \frac{2\theta(\tilde{h})}{\pi} \,\tilde{R}(\tilde{h})
   \hspace*{5mm} \text{and} \hspace*{5mm}
 \lambda_\phi = \tilde{R}_B = \tilde{R}(\tilde{h})\sin(\theta(\tilde{h})),
\label{eqn:spherical-strain_app}
\end{align}
where $\tilde{R}(\tilde{h})$ is the reduced radius as 
given by the volume constraint  \eqref{eqn:cap-volume-constraint_app} and 
$\theta(\tilde{h})$
is the opening angle of the spherical caps according to 
\eqref{eqn:wetting-angle_app}.
Obviously $\lambda_s< \lambda_\phi$ because $\theta <\pi/2$ and 
$R > R\sin\theta$,
i.e., the stretches (\ref{eqn:spherical-strains}) are anisotropic. 
Inserting these stretches  into the constitutive
relations \eqref{eqn:constitutive-laws} we also obtain 
anisotropic  stresses, $\tau_s <\tau_\phi$. 
Therefore, assuming
constant tensions in a spherical cap formed from a hemispherical rest shape 
violates the
force balance conditions \eqref{eqn:equilibrium_var}.  
For $\gamma \geq Y_\mathrm{2D}$, i.e., small deviations from the 
fluid-like behavior
deviations from force balance are also small, and we can use
spherical caps as an approximation. 

Inserting the  uniform strains
\eqref{eqn:spherical-strains}  into the constitutive
relations \eqref{eqn:constitutive-laws} gives the stresses
$\tau_s(\tilde{h})$ and $\tau_\phi(\tilde{h})$ as 
a function of the reduced spherical cap height $\tilde{h}$.
Then we employ the Neumann condition 
$f_r=0$ from eq.\ \eqref{eqn:young-equation} in the 
symmetric case, i.e., $\theta = \psi^-(\ell)  = \pi - \psi^+(\ell)$ and
$\gamma=\gamma_\mathrm{A} = \gamma_\mathrm{B}$, 
 the geometric relation (\ref{eqn:wetting-angle_app})
and the volume constraint \eqref{eqn:cap-volume-constraint_app}, to obtain 
\begin{align}
\sigma &= 2\tau_s(\tilde{h})\cos\theta(\tilde{h})
    = 2\tau_s(\tilde{h})\left(1-\frac{\tilde{h}}{\tilde{R}(\tilde{h})}\right)
= 4\tau_s(\tilde{h}) \frac{1-\tilde{h}^3}{2+\tilde{h}^3} 
\label{eqn:cap-theory_app}
\end{align}
analogously to eq.\ (\ref{eqn:Neumann_app}).
Solving this equation for $\tilde{h}$ gives the height 
as a function of $\sigma/\gamma$ and $\gamma/Y_\mathrm{2D}$. 
In the fluid regime $\gamma \gg  Y_\mathrm{2D}$ ($\tilde\gamma \gg 1$ 
in Figs.\ \ref{fig:wetting_exponent}(A,B)), we have 
 $\tau_s(\tilde{h})\approx \gamma$ and find 
 $\cos\theta \approx  2(1-\tilde{h}) \sim \sigma/\gamma$ for 
small $\sigma/\gamma$. 
In the elastic regime $\gamma \ll  Y_\mathrm{2D}$ ($\tilde\gamma \ll 1$ 
in Figs.\ \ref{fig:wetting_exponent}(A,B)), we have 
$\tau_s(\tilde{h}) \sim  Y_\mathrm{2D}(1-\tilde{h})^2$, 
which gives $\cos\theta \approx 2(1-\tilde{h}) 
\sim (\sigma /Y_\mathrm{2D})^{1/3}$ for small $\sigma/Y_\mathrm{2D}$. 
From the Laplace-Young equation (third eq.\ in 
\eqref{eqn:thin-shape-equations}), we then find the pressure $p_0$ as
 a function of $h$,
\begin{align}
p_0(\tilde{h}) &= \frac{3\tilde{h}^2(\tau_s(\tilde{h})+\tau_\phi(\tilde{h}))}
   {2+\tilde{h}^3},
\end{align}
and, thus, as a  function of $\sigma/\gamma$ and $\gamma/Y_\mathrm{2D}$. 

Note that instead of solving \eqref{eqn:cap-theory_app},
we could likewise minimize the free energy for the elastic lens,
which is obtained analogously to \eqref{eqn:free-energy-lens}, but 
with an additional elastic energy term as in \eqref{eqn:hooke-energy}.

\subsection{Pogorelov approximation for finite $E_\mathrm{B}$}
\label{sec:Pogorelov_app}

For $E_\mathrm{B}>0$ the edge of the elastic lens becomes 
rounded as shown in  Fig.\ \ref{fig:spherical-caps-pogorelov}(A).
The curvature $\kappa_s$ at the AB-interface is a characteristic 
geometric feature of these shapes, which depends on the elastic 
properties of the capsule and the interface load. 
In Fig.\ \ref{fig:wetting_exponent}(C),
we found a scaling $\kappa_s \sim
E_\mathrm{B}^{-1/2}$ for the curvature $\kappa_s$  from
numerical simulations. 
We can derive this scaling from 
an adaption of Pogorelov's theory
\cite{pogorelov1988bendings} as it has been used in
Ref.\ \citenum{knoche2014a,knoche2014b}.
The reader is referred to Fig.\ \ref{fig:spherical-caps-pogorelov}(B)
for the involved geometric quantities.
We start from an elastic lens shape for 
vanishing bending rigidity $E_\mathrm{B} \approx 0$,
which is described by the contour
$(r(s_0), z(s_0))$ and has 
a sharp edge at the AB-interface. If the bending rigidity 
is introduced, the shape  becomes rounded at the edge, 
which is described by an additional displacement field 
 $(v(s_0), u(s_0))$
such that 
\begin{align}
(r(s_0), z(s_0)) \rightarrow (r(s_0)+v(s_0), z(s_0)+u(s_0)).
\end{align}
during rounding.  
Rounding involves two additional energies contributing to the shape:
the additional bending energy $U_B$ associated with the bending rigidity 
and an additional mechanical work $U_\sigma$ performed 
against that the surface tension 
$\sigma$ because rounding displaces the 
three phase contact line against the surface tension $\sigma$.

Since the interface load acts in radial direction, it is
reasonable to assume that $u(s_0)=0$, i.e., only radial shape perturbations
occur. This allows us to write the total  energy change, induced by 
bending and pulling against $\sigma$ 
 while starting at the elastic lens, as
\begin{align}
\begin{split}
U &= U_B + U_\sigma \\
  &= 2\int_{\ell}^{\ell+\epsilon}ds_0 
    \left[\pi R_D E_\mathrm{B} v''(s_0)^2 
   - \pi R_D \sigma v(s_0) \delta (s_0-\ell)\right]
\end{split}
\label{eqn:pogorelov}
\end{align}
where $\epsilon$ describes a small arc length 
region over which the kink of the shape is
rounded and $R_D$ is the radius of the circle in the interface plane, i.e.,
the radius of the interface cross-section. 
 In eq.\ \eqref{eqn:pogorelov} we neglect the apparent
hoop stretching as well as higher order and constant terms of the bending
energy (for a detailed analysis, see Ref.\ \citenum{pogorelov1988bendings}).
Throughout the following calculations we use approximations for small
$\alpha$, 
which is the turning angle of the shape at the AB-interface and related
to the opening angle of the elastic lens via $\alpha = \pi/2 - \theta$. 
To find the corresponding characteristic arc length scale $\xi \sim
\alpha\kappa_s^{-1}$ on which the shape gets bent at the interface, we
non-dimensionalize eq.\ \eqref{eqn:pogorelov} according to
\begin{align}
\begin{split}
s_0 &= \xi \bar{s}_0, \qquad
-v(s_0) = \xi\alpha \bar{v}(\bar{s}_0), \\ 
-v''(s_0) &= \frac{d^2 v(s_0)}{ds_0^2} 
  = \alpha\,\frac{\xi d^2 \bar{v}(\bar{s}_0)}{\xi^2 d\bar{s}_0^2} 
    = \frac{\alpha}{\xi}\,\bar{v}(\bar{s}_0)'',
\end{split}\label{eqn:pogorelov2}
\end{align}
and recast the energies in 
 \eqref{eqn:pogorelov} as
\begin{align}
U_B &=   E_\mathrm{B}R_D \frac{\alpha^2}{\xi}
     \int_{\bar\ell}^{\bar\ell+\bar\epsilon}d\bar{s}_0 \bar{v}(\bar{s}_0)''^2 ,
\\
U_\sigma &= \pi R_D\sigma \xi \alpha
     \int_{\bar\ell}^{\bar\ell+\bar\epsilon}d\bar{s}_0
     \bar{v}(\bar{s}_0)\delta(\bar{s}_0-\bar{\ell}).
\label{eqn:min-pogorelov}
\end{align}
Minimizing the total energy $U= U_B+U_\sigma$ with respect to $\xi$ 
we find  the scaling relations
\begin{align}
\xi \sim \alpha^{1/2} \sqrt{\frac{E_\mathrm{B}}{\sigma}},\qquad
U = U_0 \sim R_D \alpha^{3/2} \sqrt{E_\mathrm{B}\sigma}.
\label{eqn:xi-scaling}
\end{align}
The dimensionless integrals in eq.\  \eqref{eqn:min-pogorelov}
only contribute  numerical prefactors; the essential result is 
the scaling behavior of  $U_0$ and $\xi$.
 Note that the above scaling relations still depend on $\alpha$.
This dependence can be eliminated by employing the symmetric Neumann force 
balance condition \eqref{eqn:cap-theory_app}, 
$\sigma = 2\tau_s(\tilde{h})\sin\alpha$ and the geometric 
relation $\sin\alpha  =  2(1-\tilde{h}^3)(2+\tilde{h}^3)$.
In the fluid regime $\gamma \gg Y_\mathrm{2D}$, we have 
 $\tau_s(\tilde{h})\approx \gamma$ and find $\sin\alpha \sim \sigma/2\gamma$.
In the elastic regime $\gamma \ll Y_\mathrm{2D}$, we have 
$\tau_s(\tilde{h}) \sim  Y_\mathrm{2D}(1-\tilde{h})^2\sim 
Y_\mathrm{2D} \sin^2\alpha$ for small $\alpha$. 
For small $\alpha$, this leads to 
\begin{align}
\alpha &\sim \begin{cases} 
       (\sigma/Y_\mathrm{2D})^{1/3}, & \gamma \ll Y_\mathrm{2D}
     \hspace*{5mm} \text{(elastic)} \\
       \sigma/\gamma, & \gamma \gg Y_\mathrm{2D} 
     \hspace*{5mm} \text{(fluid)}, \end{cases}
\label{eqn:alpha-scaling}
\end{align}
in the two regimes. The above scaling behavior of
$\alpha$ is verified numerically in 
Fig.\ \ref{fig:spherical-caps-pogorelov}(C),
where we measured $\alpha$ for an elastic lens
dependent on $\sigma / Y_\mathrm{2D}$ for two different
values of $\gamma$ corresponding to the elastic respectively
fluid regime. Inserting these scaling relations
into eq.\  \eqref{eqn:xi-scaling} gives
\begin{align}
\xi &\sim \begin{cases} 
       E_\mathrm{B}^{1/2}\sigma^{-1/3}\,Y_\mathrm{2D}^{-1/6}, 
            & \gamma \ll Y_\mathrm{2D}
      \hspace*{5mm} \text{(elastic)} \\
       E_\mathrm{B}^{1/2}\gamma^{-1/2}. & \gamma \gg Y_\mathrm{2D}
       \hspace*{5mm} \text{(fluid)}. \end{cases}
\end{align}
From $\kappa_s \sim \alpha / \xi$ we finally obtain the scaling laws 
for the shell's curvature at the AB-interface
\begin{align}
\kappa_s &\sim \begin{cases}
  E_\mathrm{B}^{-1/2}\sigma^{2/3}\,Y_\mathrm{2D}^{-1/6},
       & \gamma \ll Y_\mathrm{2D}
         \hspace*{5mm} \text{(elastic)} \\
                   E_\mathrm{B}^{-1/2}\sigma \gamma^{-1/2}, & 
       \gamma \gg Y_\mathrm{2D} \hspace*{5mm} \text{(fluid)}.
\end{cases}
\label{eqn:kappa-scaling}
\end{align}
The dependence on  the bending modulus,
 $\kappa_s \propto E_\mathrm{B}^{-1/2}$, is universal, i.e.,
independent of whether we are in the fluid or elastic  regime. 
For comparison, spherical shells which buckle upon deflation have 
$\kappa_s \propto E_\mathrm{B}^{-1/4}$ at the spherical 
rim of the indentation \cite{knoche2014a,knoche2014b}.

\end{document}